# Thermodynamic and Kinetic Tests for Low-Pressure Metal-Flux Growth of Diamond, with an Extension to Cubic Boron Nitride

**Rodney S. Ruoff**[1,2,3,4,*]

[1] Center for Multidimensional Carbon Materials (CMCM), Institute for Basic Science (IBS), Ulsan 44919, Republic of Korea
[2] Department of Chemistry, UNIST, Ulsan 44919, Republic of Korea
[3] Department of Materials Science and Engineering, UNIST, Ulsan 44919, Republic of Korea
[4] School of Energy and Chemical Engineering, UNIST, Ulsan 44919, Republic of Korea
[*] Corresponding author: ruofflab@gmail.com

## Conspectus

In 2024, we reported diamond growth in a liquid gallium-iron-nickel-silicon (Ga-Fe-Ni-Si) alloy at about 1 atm, where bulk graphite is thermodynamically more stable than bulk diamond.[1] Isotope labeling showed that methane supplied carbon to the diamond and, in the tested crucible-contact configurations, that a graphite crucible supplied additional carbon. Postmortem cross-sections located recovered graphite mainly at the diamond-metal interface. These observations raised a specific question: under what reservoir and interfacial conditions can a diamond front continue to advance even though graphite remains the stable bulk carbon phase? I address that question by examining the smallest event that can repeat without limit: adding one carbon atom at a kink on a growing diamond step. A front can advance indefinitely only if that event leaves the front unchanged apart from a lattice translation, and that requirement fixes what the surrounding liquid must supply.

Until that work, growing diamond from a metal melt meant working at high pressure. Here the diamond formed near 1 atm, in a cooler, carbon-rich region below the surface of the melt. Bulk graphite is still the stable carbon phase under those conditions, and nothing in the experiment changed that. The experiment did give us a liquid reservoir whose temperature and carbon content can be set and measured, and with it a way to ask why diamond grew instead of graphite.

I consider a broader experiment: a diamond seed already in place, in contact with a carbon-bearing metal flux. Here, “metal flux” means a metal or alloy that is liquid under the growth conditions and serves as the carbon-bearing reservoir. The Ga-Fe-Ni-Si system under methane/hydrogen feeding is the demonstrated reservoir system, but the thermodynamic accounting is not specific to this alloy. Whether it applies to another flux must be tested.

I describe one specified front transition (the front is the interfacial region where the growing crystal meets the liquid) using the finite-difference transfer free energy $\Delta Y$. The initial and final states must each be a constrained-equilibrium front basin, or a metastable front basin that persists long enough to be prepared again. A single molecular snapshot will not serve, because it has no reversible free energy. $\Delta Y$ compares their reversible free energies and subtracts the chemical work (the $\mu \cdot dN$ work of matter transfer) supplied by the reservoirs. In words, $\Delta Y$ is the free-energy change of the front after subtracting the chemical potential of every species transferred from its reservoir. A negative $\Delta Y$ makes the specified event thermodynamically downhill; $\Delta Y = 0$ marks reversible balance with the exact reverse event; a positive $\Delta Y$ favors the reverse event. The sign gives the thermodynamic direction of the specified event and nothing more—not the barrier, not the rate, and not which product is retained.

Everything below rests on one event: a completed attachment at a kink. This is the only event that can repeat without limit, which is why it fixes what the liquid must supply. One complete growth unit is incorporated, the kink moves by one lattice repeat, and the final front is equivalent to the initial front. Surface terms therefore cancel and the added unit joins the interior, so in the macroscopic limit set out in Section 2 the reversible cost of that repeat equals the bulk free energy per growth unit. Consequently, the carbon activity required to grow macroscopic diamond always exceeds the activity at which bulk graphite is saturated. On a graphite-normalized activity scale, bulk graphite is at $S_G = 1$ and the pure-diamond equilibrium threshold on that scale is $S_{G,D}^{eq} = \exp[(g_D^{bulk} - g_G^{bulk})/RT]$, about 2.0 at 1300 K with the thermochemical data and model adopted here.

This bulk constraint does not determine which phase is observed. Graphite nucleation may be slow, diamond attachment may be fast, and finite graphite states may have reversible boundaries different from that of bulk graphite. These possibilities must be tested separately. A controlled finite graphite object provides a pairwise comparison; failure to detect spontaneous graphite during a finite run provides only a kinetic rate bound.

The chemical-potential argument holds for any flux. The numerical values must be determined again for each one, but each flux maps composition and feed onto carbon activity differently, and each has its own barriers, transport, and competing sinks. Six tests follow: (1) calibrate $S_G = 1$ and validate the activity map above saturation; (2) bracket reversible retreat and advance of a registered diamond front; (3) demonstrate isotope-attributed outward overgrowth; (4) measure the reversible boundary of one specified finite competitor; (5) quantify stochastic competitor rates and detection efficiency; and (6) close the phase-resolved carbon balance and only then attempt a thin-film or other scale-up geometry (SI Protocol 6).

The same accounting can be extended to boron nitride, but not with a single chemical-potential coordinate. BN growth requires separate control of the BN-pair chemical potential and the B/N imbalance, plus the electron electrochemical potential when the declared event exchanges electrons with a reservoir. These coordinates define compatibility conditions; they do not by themselves establish preferential growth of cubic BN (cBN) over layered BN.

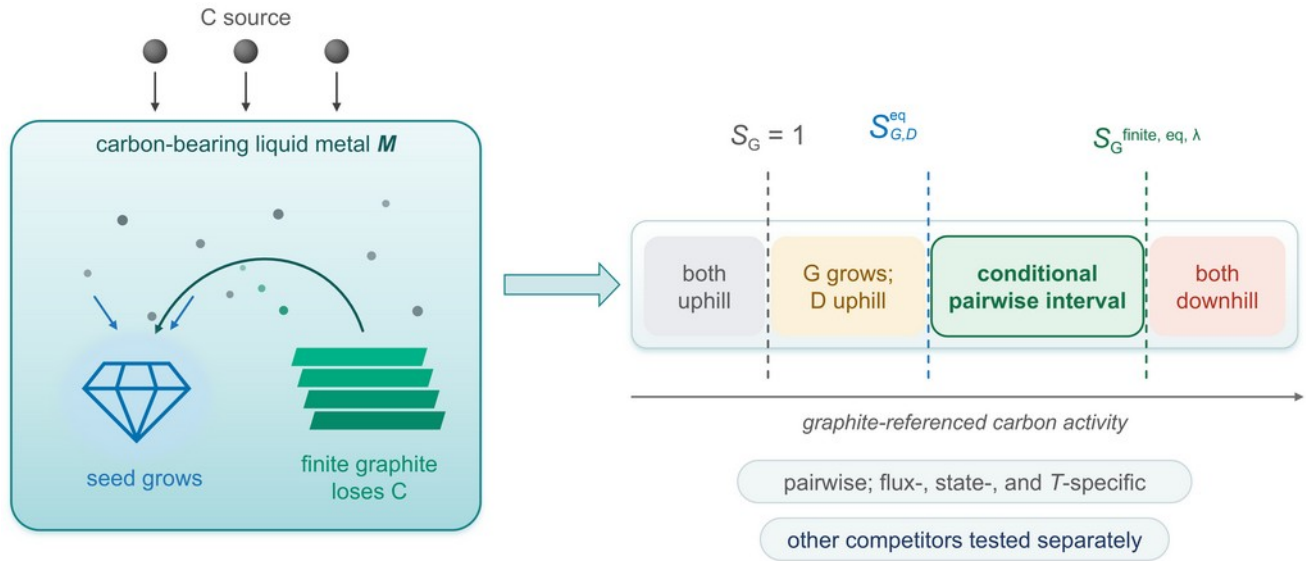

***Conspectus graphic.*** *For a diamond seed in a chosen carbon-bearing flux, the proposed operating interval compares two measured boundaries: the diamond repeat boundary and the reversible boundary of one specified finite graphite state. This comparison does not exclude other graphite states, carbides, amorphous carbon, or stochastic nucleation.*

## 1. Demonstrated growth and the unresolved selection problem

Our metal-flux growth program established that diamond could nucleate and grow at approximately 1,025 °C and approximately atmospheric pressure in a Ga-Fe-Ni-Si melt fed by methane and hydrogen. The products formed in the cooler part of a carbon-rich subsurface region. The quoted temperature is the coldest probed position in a melt spanning roughly 60 °C laterally at the growth depth, so the demonstrated reservoir is not isothermal, and a single flux temperature is an approximation. Substitution of $^{13}CH_4$ produced isotope-labeled diamond, and the observations implicated Si in the formation or stabilization of tetrahedrally bonded carbon clusters. In the tested crucible-contact configurations, isotope results support the graphite crucible as the dominant additional $^{12}C$ source. Postmortem cross-sectional TEM located recovered graphite mainly at the diamond-metal interface. These observations establish low-pressure diamond growth and carbon transfer through the metal flux, but they do not establish when the interfacial graphite nucleated, its mechanistic role, or why diamond rather than graphite was retained.[1]

With 99 at% $^{13}CH_4$ feed, growth in direct contact with the graphite crucible was $^{12}C$-rich, growth isolated from the crucible by dense graphite-like plates was purely $^{13}C$, and partial isolation gave an intermediate result. This contrast makes the metal feedstocks and adventitious carbon unlikely to be the dominant $^{12}C$ sources in those configurations, but it does not exclude minor sources or isotope-fractionation effects. The plates did not contribute at the measurement sensitivity. Their composition, density, and degree of graphitization were not fully characterized, so the result does not establish selectivity for a particular graphite state. The methane and crucible contributions remain semi-quantitative because they were inferred from Raman intensities.[1]

Recovered graphite was spatially heterogeneous. Its interfacial location is consistent with formation at the diamond-metal boundary, but it does not determine the sequence of diamond and graphite nucleation or the path by which carbon reached either phase. Regions without detected graphite can reflect stochastic nucleation, uncontrolled heterogeneity, or limited detection and recovery. A null result bounds an event rate only when the event, exposure, detection and recovery efficiency, stochastic model, and confidence level are specified.

The experimental trajectory began with a non-peer-reviewed seeded Ga-Si study that reported homoepitaxial diamond pyramids and isotope transfer from methane and the seed.[2] Simulations then proposed carbon-chain and ring-forming pathways in liquid Ga,[3] and a subsequent Viewpoint outlined prospective ways to deliver C, B, and N into liquid metals with low reported equilibrium solubility.[4] These studies motivate the present analysis, but none establishes a universal flux, mechanism, or operating window.

I consider a pre-existing diamond seed in contact with a chosen carbon-bearing metal flux, denoted $M$. Here, the flux is a metal or alloy that is liquid under the growth conditions. Carbon must reach the seed-liquid interface; the reservoir components and external controls must be defined; relevant competing phases and reactions must be identifiable; and the interface must be interrogated reproducibly enough to measure advance and retreat and to account for the carbon. The seed must also remain physically and chemically viable. Each candidate $M$ requires its own relation between composition and carbon activity, dissolved-carbon speciation, interfacial states, competing carbon phases, barriers, transport coefficients, wetting, gas-liquid chemistry, and loss pathways. The accounting is the same for every candidate $M$. Each $M$ must be characterized before it can be applied.

Ga-Fe-Ni-Si with methane and hydrogen is therefore the reservoir system in which isotope-resolved diamond growth has been demonstrated. Whether another flux will serve is open. To my knowledge, the complete seeded reversible-front program proposed here has not been reported. The analysis asks three separate questions: (i) Is one specified front event thermodynamically downhill? (ii) How rapidly can it occur? (iii) Where does the carbon remain after all competing processes? The last question asks whether a product seen locally accounts for the carbon retained throughout the reactor.

Jiang et al. subsequently reported faceted micrometer-scale diamond particles at 900 °C and 1 atm using Ga-In, nanodiamond seeds, Si-containing additives or substrates, $H_2/N_2$, and ferrocene supplied upstream as a combined carbon-containing molecular precursor and Fe source.[63] This result is of interest because it extends ambient-pressure liquid-metal diamond synthesis to a different liquid composition and to a feed that is initially solid. Because the study did not report a control in which ferrocene alone was omitted, isotope tracing, or a matched comparison between carbon sources, the carbon incorporated into the reported diamond particles cannot be assigned uniquely to ferrocene. The experiment also does not distinguish the proposed carbon-supply function of ferrocene from the effect of the Fe released during its decomposition.

Carbon need not be supplied as a gas. Condensed sources—including bulk diamond, nanodiamond, glassy or other amorphous carbon, molecular carbon solids, and carbides—can differ substantially in dissolution, transport, reaction with the liquid, and competing-product formation. The bulk free energy of a source does not by itself determine the carbon potential at the seed. Carbon must enter the liquid, move through it, and reach the seed without being consumed by graphite, amorphous carbon, carbide formation, deposits elsewhere in the reactor, or volatile products. A temperature gradient may be useful because source activation and incorporation at the seed need not occur at the same temperature. Supporting Information Section S2.7 gives a provisional ordering of condensed carbon sources and identifies the measurements needed to compare them.

A $^{13}$C-enriched diamond seed provides a particularly useful way to screen these sources. The candidate feed can remain at natural isotopic abundance; outward growth derived from it should then appear as a $^{12}$C-rich diamond layer extending beyond the registered surface of the original $^{13}$C-rich seed. This arrangement avoids preparing a separate $^{13}$C-labeled form of every candidate source. The isotope profile must nevertheless be registered to the original seed surface and interpreted together with seed recession, lattice exchange, surface roughness, isotope fractionation, and every other natural-abundance carbon reservoir in the apparatus.

This distinction matters particularly when the competing phases have different bonding networks. SiC polytypes and zinc-blende or wurtzite III-V structures have related bonding networks; diamond versus graphite, and cBN versus layered BN, differ in bonding network. Polytype-control examples are useful analogies for front-mediated selection, but they do not show that a low-pressure flux reverses the macroscopic $sp^2/sp^3$ ordering. I treat diamond in full. BN is taken up in Section 6, where two additional coordinates are needed and the evidence is less complete.

The reservoir transformation used below is standard thermodynamics. I apply it to one fully specified crystal-front event, find the limit in which its cost becomes the bulk free energy per growth unit, distinguish a controlled finite competitor from one that nucleates at random, and identify measurements that could show the assignments to be wrong.

## 2. One declared event and the macroscopic carbon constraint

Compare the same interfacial region before and after one specified event under the same temperature, pressure, reservoir, mechanical, and electrical controls. The initial state $s$ and final state $s'$ must be constrained-equilibrium states or reproducible metastable basins: each state must be reproducible under the imposed controls while held within its basin. An arbitrary nonequilibrium molecular snapshot does not, by itself, possess the reversible free energy used here. The complete state description and operational relaxation tests are given in SI Section S1.1.

Let $\nu_i > 0$ denote transfer of species $i$ from its reservoir into the front. The molar reservoir-transformed free-energy change is

$$\Delta Y(s \longrightarrow s') = G_{\mathrm{front}}(s') - G_{\mathrm{front}}(s) - \sum_i \nu_i \mu_i^{\mathrm{res}}.$$

Here, $G_{\mathrm{front}}$ is the reversible free energy of the declared front region and $\mu_i^{\mathrm{res}}$ is the chemical potential of reservoir species $i$. Required mechanical, electrical, or electrochemical work belongs in the stated ensemble. Within those definitions, $\Delta Y$ is an exact finite difference. Its sign gives thermodynamic direction for that event: $\Delta Y < 0$ favors the forward event, $\Delta Y = 0$ gives zero affinity relative to the exact reverse, and $\Delta Y > 0$ favors the reverse. The sign does not specify the rate.

The reference event is a completed translational repeat at a kink: one complete growth unit moves the kink by one lattice repeat and leaves the final front basin equivalent to the initial basin after translation. I use “completed translational repeat” to exclude events that leave new area, line length, defects, reconstruction, adsorbates, strain, charge, polarity, or another persistent change. Kossel, Stranski, and Kaischew established the repeat-site lineage; Burton, Cabrera, and Frank (BCF) connected kink statistics to step motion.[5–8] The present ΔY, constrained-basin, and repeat notation are modern formalizations, not historical notation. Real incorporation can include desolvation, preliminary binding, and structural reorganization,[9,10] and matched forward and reverse rate expressions must satisfy detailed balance.[11] SI Section S1.4 gives the historical and kinetic distinctions.

For an event that incorporates $n$ complete growth units, where $n$ is the number of growth units transferred in the event,

$$\Delta Y = -n(\mu_u^{\mathrm{res}} - \mu_{k,\varphi}^{\mathrm{rep}}) + \Delta G_{\mathrm{nonrep}}.$$

Here, $\mu_u^{\mathrm{res}}$ is the stoichiometric reservoir potential per growth unit; $\mu_{k,\varphi}^{\mathrm{rep}}$ is the reversible repeat cost in phase $\varphi$; and $\Delta G_{\mathrm{nonrep}}$ contains changes that remain after the completed repeats, including new area or line length, curvature, reconstruction, adsorbates, defects, strain, charge, polarity, or configurational change. In the pure, unstressed, translationally homogeneous macroscopic limit, the long-range excess per repeat vanishes and

$$\mu_{k,\varphi}^{\mathrm{rep}} = g_\varphi^{\mathrm{bulk}}.$$

Figure 1 summarizes the declared event and this limiting identity.

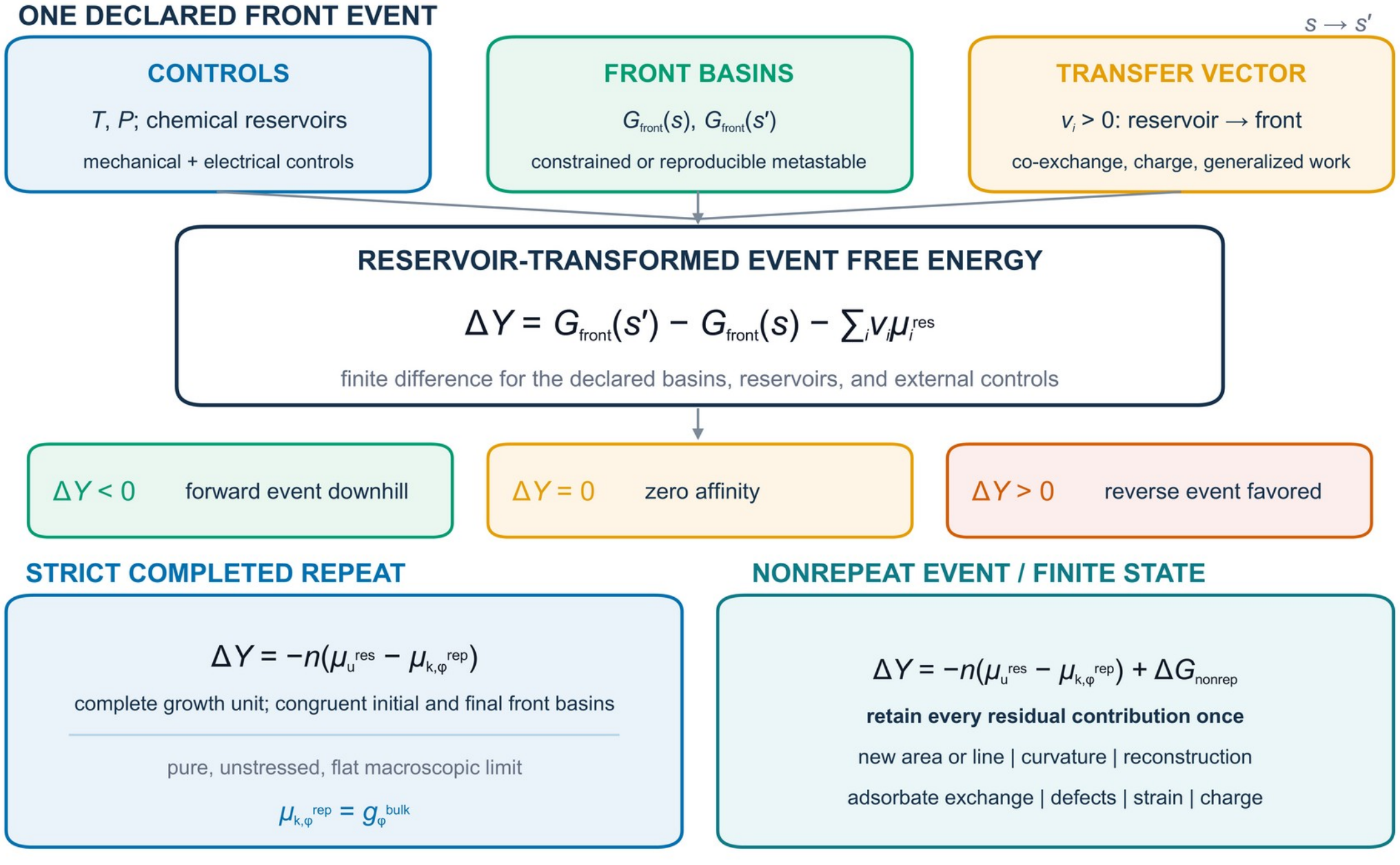

**Figure 1. From a declared front event to thermodynamic direction. The controls, front basins s and s′, and transfer vector v define the exact finite difference ΔY for the specified basins, reservoirs, and external controls. Its sign gives thermodynamic direction only. A strict completed repeat leaves congruent front basins and specializes to $\Delta Y=-n(\mu_u^{res}-\mu_{k,\varphi}^{rep})$; for a pure, unstressed macroscopic repeat, $\mu_{k,\varphi}^{rep}=g_\varphi^{bulk}$. Nonrepeat events retain every residual contribution in ΔGnonrep exactly once. Thermodynamic direction does not specify the barrier, rate, or retained inventory.**

The chosen flux can change the reservoir activity needed to reach the repeat threshold and can change every nonrepeat term, barrier, transport coefficient, wetting condition, and loss pathway. A finite interfacial benefit can support growth below the macroscopic repeat threshold only while that benefit remains. GeSi growth from Bi solution illustrates this point: interfacial-energy reduction supported about four monolayers of growth from an undersaturated solution before the layer broke into islands.[12] The example does not establish how the effect depends on the flux. It shows what happens when a finite interfacial benefit is exhausted: growth stops and the layer breaks apart.

The bulk repeat identity also does not guarantee an abundant or rapidly equilibrating kink population. Sparse or slowly replenished kinks can make kink-pair creation and propagation kinetic bottlenecks without changing the thermodynamic repeat cost. Silicon step-bunching studies show how a kink statistic and transport can define a process boundary, but their energy partition contains assumed inputs and their surface can generate new steps or facets when the existing step supply is inadequate.[13,14] Alternating-step BCF theory and diamond CVD models provide further precedents for nonequivalent step types and reaction-specific attachment sites.[15–19] These studies address how readily a growth event occurs. They do not change its repeat free energy. Their equations and limitations are in SI Sections S1.3–S1.5.

To compare fluxes without confusing composition with chemical potential, I reference the carbon chemical potential to macroscopic graphite:

$$S_G \equiv (a_C)/(a_C^{sat,G}) = \exp[(\mu_C^L - g_G^{bulk})/(RT)].$$

$S_G = 1$ is bulk graphite saturation. This is an activity scale, not a mole-fraction scale. The mapping from composition, gas feed, and temperature to $S_G$ is specific to $M$ and must be calibrated or computed with stated uncertainty. A single elemental $\mu_C^L$ is adequate only when dissolved carbon-containing species interconvert rapidly enough to share local chemical equilibrium; otherwise, species-specific affinities are required.

The pure macroscopic diamond threshold on this graphite-normalized scale is

$$S_{G,D}^{eq}(T,P)\equiv\exp[(g_D^{bulk}-g_G^{bulk})/(RT)].$$

At the pressures and temperatures considered here, $S_{G,D}^{eq} > 1$. For $S_G < 1$, incorporation into both bulk graphite and pure macroscopic diamond is uphill. For $1 < S_G < S_{G,D}^{eq}$, graphite incorporation is downhill, but the diamond repeat remains uphill. For $S_G > S_{G,D}^{eq}$, completed-repeat incorporation into both phases is downhill. There is therefore no macroscopic activity interval in which pure diamond-repeat growth is downhill while bulk graphite-repeat growth is uphill.

White *et al.* measured the ambient diamond-graphite gap as 3.170 ± 0.150 kJ mol$^{-1}$, with graphite lower.[20] Combining that anchor with the temperature increment from the Fried-Howard carbon equation of state[21] gives approximately 7.23 kJ mol$^{-1}$ and $S_{G,D}^{eq} \approx 1.95$ at 1300 K. Two descriptions that do not share the reference-state entropy of that equation of state place the threshold nearer 2.0: integrating assessed heat-capacity functions for graphite and diamond gives 1.98, and re-evaluating the same increment with the accepted standard entropy of diamond gives 2.01 (SI Section S2.2). I therefore use "about 2.0 at 1300 K with the stated thermochemical data and model." The descriptions examined span approximately 1.95 to 2.01; uncertainty beyond that spread is not quantified. The chosen flux can change the composition corresponding to this threshold and every relevant barrier, transport coefficient, or loss pathway, but it cannot lower the pure, unstressed macroscopic diamond threshold below bulk graphite saturation. Figure 2 separates bulk stability, repeat-event direction, and kinetic barriers.

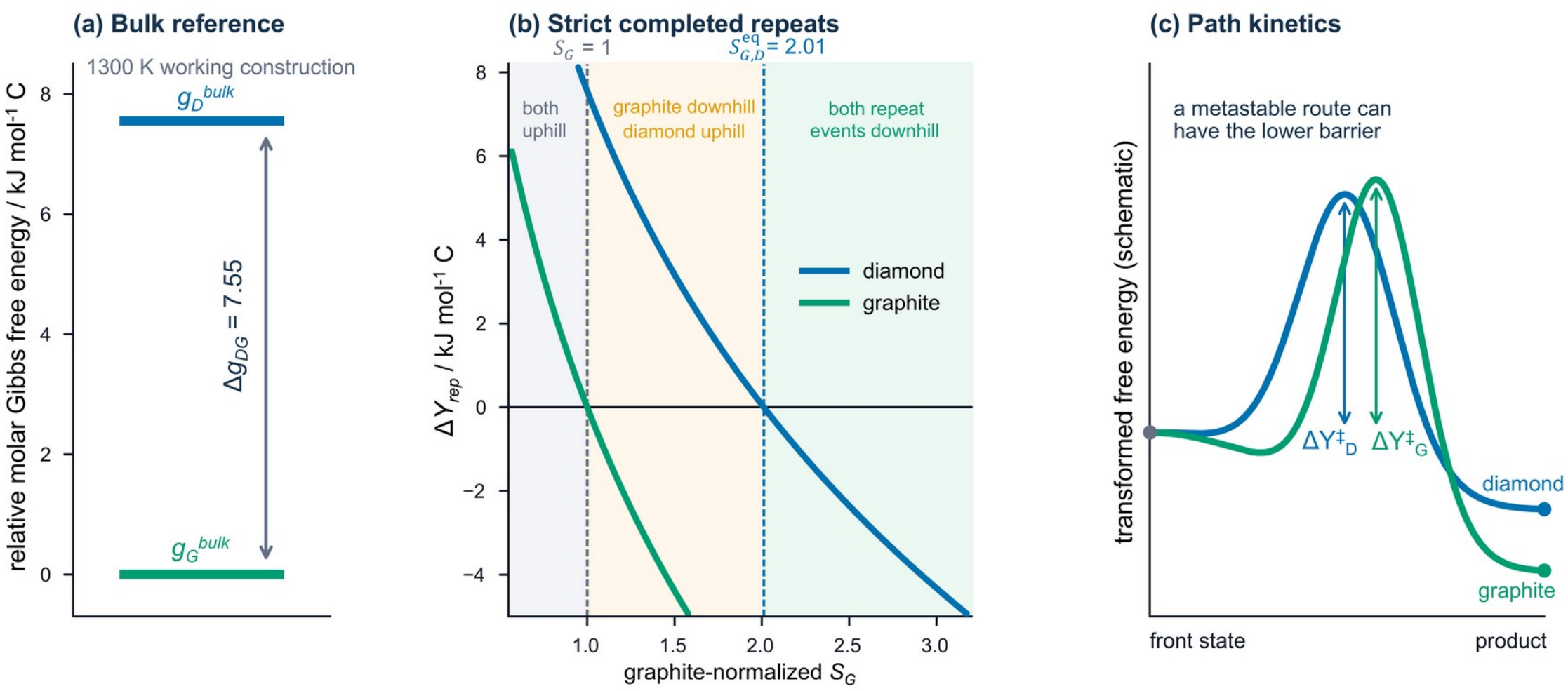

**Figure 2. Bulk ranking, repeat-event direction, and kinetic barriers are distinct.** (a) Graphite has the lower macroscopic molar Gibbs energy near 1 atm; the plotted gap uses the 1300 K thermochemical data and model stated in Section 2. (b) The exact completed-repeat relations are $\Delta Y_G^{rep}=-RT\ln S_G$ and $\Delta Y_D^{rep}=RT\ln(S_{G,D}^{eq}/S_G)$; both are downhill only for $S_G>S_{G,D}^{eq}$. (c) Observed pathways additionally depend on reaction-coordinate-dependent barriers, prefactors, transport, available sites, losses, and observation time. Barrier curves are schematic.

## 3. Finite competitors, kinetic access, and retained product

The bulk constraint does not settle every finite comparison. Edge, support, termination, reconstruction, strain, adsorbates, and interfacial contributions can change the marginal carbon-transfer cost of a finite graphite state even though they vanish per atom in the bulk limit. The specified state must include its layer number or thickness, edge structure, support, strain, defects, passivation, interfacial state, and allowed reservoir exchanges. Its reversible boundary must be measured or calculated along a declared growth or dissolution path. SI Sections S2.3–S2.6 give the full construction and include carbide competitors.

Let χG denote the declared graphite state and λ its growth or dissolution path at stated flux, temperature, pressure, support, and external controls. Here, $N_C$ is the amount of carbon in the finite graphite state, expressed in moles of carbon on the molar basis used here. The marginal carbon-transfer cost is a total derivative of a partially reservoir-transformed free energy in which carbon remains untransformed:

$$\mu_{C,G}^{\text{finite},\lambda}=(dY_G^{(C)}[N_C,\chi_G^{\lambda}(N_C),M])/(dN_C).$$

This derivative follows structural relaxation and any exchange of non-carbon species allowed along $\chi_G^{\lambda}(N_C)$. A partial derivative at fixed $\chi_G$ is appropriate only for a genuinely fixed state family. A strongly carbide-forming flux requires a corresponding metal chemical potential and cannot be bounded on the graphite-normalized coordinate alone.

Seeded diamond is thermodynamically favored over one declared finite graphite path only if

$$\mu_{C,G}^{\mathrm{finite},\lambda} > g_D^{\mathrm{bulk}},$$

Equivalently, a pairwise interval on the graphite-normalized scale requires

$$S_{G,D}^{\mathrm{eq}} < S_G^{\mathrm{op}} < S_G^{\mathrm{finite,eq},\lambda}(N_C,\chi_G,M).$$

The upper boundary belongs to the specified object alone. Graphite in general has no single boundary of this kind. The interval may be narrow, inaccessible, or absent. Even if it exists, it does not suppress other graphite states, amorphous carbon, carbides, bulk graphite nuclei, or wall deposits. Those require separate state and rate bounds. The interval cannot be inferred from failure to observe spontaneous graphite during a finite run.

Large-scale atomistic calculations and modern potential-training methods could be used to screen candidate finite graphite states.[22,23] A trained model, however, is useful here only if it has been validated for the chosen liquid, the relevant interfaces, and the competing carbon states. Its uncertainty must be smaller than the approximately 0.078 eV per carbon atom diamond-graphite separation used here. The continuum edge estimate gives the scale of one contribution: at $\ell = 5$ nm and $\beta = 1$ eV nm−1, the marginal edge term is approximately 0.25 kJ mol−1 C, about 30 times smaller than the 7.55 kJ mol−1 working gap. Matching the gap by this edge term alone would require dimensions at which the continuum model is no longer quantitative.[24] At that scale, atom-specific effects—including edge reconstruction, chemical termination, interaction with the supporting metal or substrate, passivation, strain, and the integer number of graphene layers—could matter. None has yet been shown to provide the required shift.[25–28]

For an isotropic self-similar two-dimensional feature with edge energy $E_{\mathrm{edge}} = \beta\mathcal{P}$ and area $A = N_C\bar{A}_C$, the marginal edge contribution is

$$\delta\mu_{\mathrm{edge}} = (\beta\bar{A}_C\mathcal{P})/(2A) = (\beta\bar{A}_C)/(2\ell), \qquad \ell \equiv A/\mathcal{P}.$$

The factor of one-half distinguishes the marginal boundary from the mean edge excess. Here, $\mathcal{P}$ is perimeter length, $\bar{A}_C$ is basal-plane area per mole of carbon, and $\ell = A/\mathcal{P}$ is not a feature width. The estimate covers only the continuum edge term written above. It places no bound on atomistic reconstruction or on chemistry specific to the metal. Figure 3 compares the corrected marginal-edge estimate with the two finite-object tests.

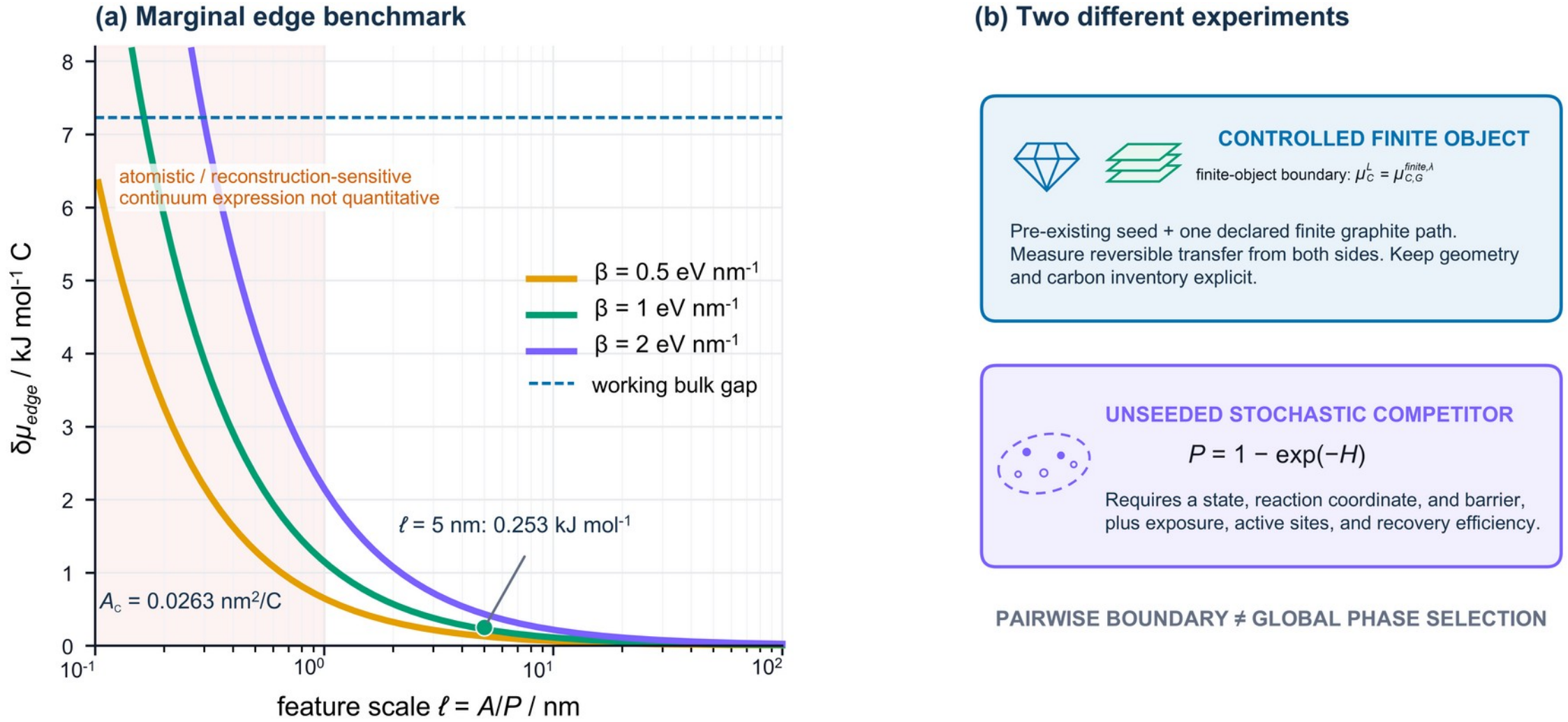

Figure 3. Corrected marginal edge benchmark and two comparison objects. (a) For a self-similar two-dimensional graphite feature, the curves use illustrative β values and are not fitted data or uncertainty bounds. At $\ell = 5$ nm and $\beta = 1$ eV nm$^{-1}$, the marginal edge

correction is approximately 0.25 kJ $mol^{-1}$ C, far below the working bulk gap. (b) A controlled finite graphite path has the general reversible boundary $\mu_C^L = \mu_{C,G}^{finite,\lambda}$. Equality with $g_D^{bulk}$ occurs only at the special condition where that finite-object boundary coincides with the diamond repeat threshold. An unseeded competitor is instead governed by its nucleation barrier, prefactor, exposure, and detection efficiency.

Thermodynamic direction, kinetic access, and retained product are separate questions. For a pathway $\alpha$, a barrier $\Delta Y_\alpha^\ddagger$ must be defined along a reaction coordinate connecting the same initial and final basins. A schematic activated rate density is

$$\mathcal{J}_\alpha=\mathcal{J}_{0,\alpha}\exp(-(\Delta Y_\alpha^\ddagger)/(RT)).$$

The prefactor includes the event exposure basis, available sites, concentrations, transport, and transmission coefficient. Solution-growth measurements demonstrate that a condensed reservoir can contribute substantially to an incorporation barrier, but the reported magnitude cannot be transferred to metal-flux diamond growth without measurement.[29] A lower final-state free energy does not imply a lower barrier, and a lower barrier does not guarantee a retained product after loss and transformation. Metastable persistence and Ostwald-type sequences depend on both state free energies and connecting barriers.[30,31]

For the specified pathway α, the molar exponent $\Delta Y_\alpha^\ddagger/RT$ and the corresponding per-event exponent $\Delta G_{event,\alpha}^\ddagger/k_B T$ describe the same barrier crossing when their event bases are matched. An Arrhenius slope obtained while temperature also changes carbon activity, speciation, transport, available sites, or the front state is only an apparent activation energy. It cannot by itself be assigned to a unique microscopic barrier.

Under a specified monitoring and recovery protocol, a null result bounds the rate of the defined graphite event only when the exposure, detection and recovery efficiency, stochastic model, and confidence level are stated. It does not locate an equilibrium boundary. Likewise, the observed point at which growth of a facet appears to stop need not be that facet's bulk equilibrium point. Pure macroscopic diamond facets share one intrinsic repeat threshold but can differ in step density, barriers, passivation, stress, curvature, and mobility. Diamond dissolution and step-flow studies show that interface retreat and facet-dependent kinetics can be measured.[32–34] In situ confocal laser scanning microscopy of a SiC seed in Si-Ni and Si-Cr solvents tracks individual nanoscale steps through a transparent seed at 1600–1800 °C and has recorded a front that reversed from advance to retreat and back within one cooling transient.[35,36] That reversal followed an uncontrolled temperature ramp, not a calibrated carbon potential, so it illustrates the finite-time zero crossing of Figure 4b rather than a located boundary. SiC is $sp^3$-bonded throughout, so these observations bear on the measurement and not on $sp^2/sp^3$ selection.

The retained-diamond inventory must finally include attachment, detachment, etching, graphitization, and other losses on a common basis of carbon amount, interfacial area, and time. High intrinsic attachment can coexist with little retained diamond, whereas a high diamond fraction among recovered solids can coexist with low feed yield if most carbon remains dissolved, leaves the reactor, or deposits elsewhere. SI Section S3 gives the rate, stochastic-detection, and common-basis inventory equations. Figure 4 shows how kinetic and experimental effects can shift the observed point at which diamond growth appears to stop away from the thermodynamic threshold.

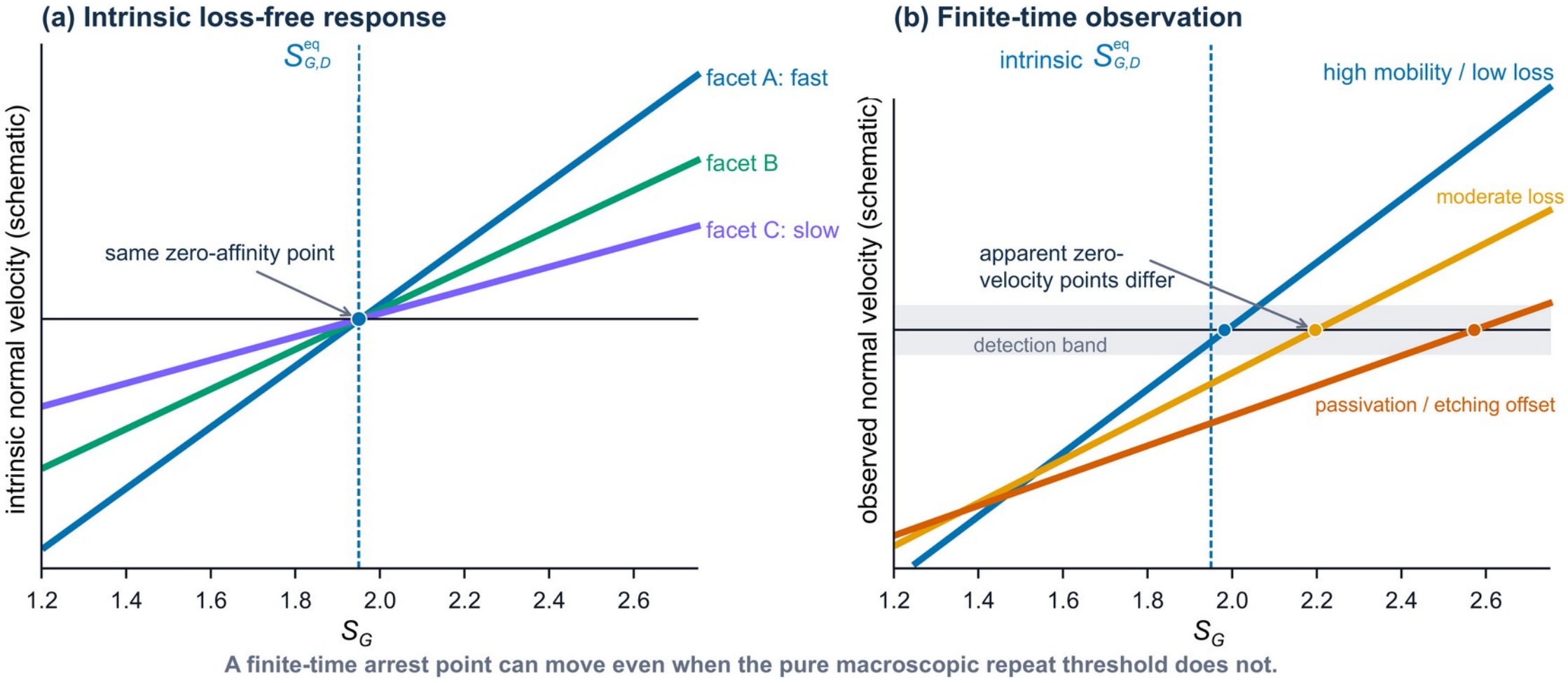

**Figure 4. One intrinsic diamond repeat threshold; finite experiments can show shifted apparent arrest.** (a) Pure, unstressed, translationally homogeneous macroscopic facets share the same completed-repeat zero-affinity threshold $S_{G,D}^{eq}$, although their mobilities differ. (b) Independent losses, passivation, etching, transport, and detection limits shift observed velocity curves and their apparent zero crossings. A finite-time zero-velocity locus therefore need not equal the intrinsic thermodynamic threshold. Curves and facet ordering are schematic.

## 4. What isotope and interface measurements establish

Isotope labeling can identify where the carbon came from only when the source composition, mixing, isotope fractionation, spatial registration, and measurement uncertainty are known. No detectable change in mass or height establishes only that the net change remained below the stated detection limit during the measurement. Attachment and removal could still have occurred at equal rates. Isotope turnover without a net dimensional change supports active lattice exchange only after contamination, unresolved diffusion, depth-response effects, and sampling or recovery artifacts have been excluded.

Three observations build the claim in sequence. Isotopically enriched carbon in confirmed diamond shows that carbon from the labeled source entered diamond. Crystallographic continuity shows that the new diamond is connected to the seed rather than being a separate crystallite. Registered displacement beyond the original interface establishes net outward growth. Together, these observations establish source-attributed overgrowth, but they do not locate a thermodynamic boundary. Locating that boundary requires reversible retreat and advance under calibrated reservoir conditions.

Isotope-labeled diamond establishes that reservoir carbon entered $sp^3$-bonded diamond, but not necessarily that the interface advanced beyond its original position. Net reservoir-fed overgrowth requires isotope attribution, crystallographic continuity with the seed, and registered outward interface motion. Buried markers, protected reference regions, or repeated step maps tied to durable fiducials outside the reactive region can provide that registration. Quench controls, residue removal, competitor identification, and phase-resolved carbon and isotope balances are also required. Carbon-isotope fractionation and depth profiling are measurable,[37–40] but each observation supports only the claim it directly tests. SI Section S3.4 and Table S4 state the full evidence ladder.

## 5. An experimental program that can falsify the assignments

The logic is general, but the experiment is specific to each system. A macroscopic graphite coupon defines $S_G = 1$ only in an equilibrium calibration mode. Stationary mass in an open gas-fed reactor can instead represent a nonequilibrium steady state with carbon flowing through the coupon. A valid equilibrium reference requires isolation of throughput, reversible convergence from carbon-poor and carbon-rich sides, and independence from flow and coupon area after equilibration—or an independently validated carbon-potential measurement.

For a suitably chosen metal flux in contact with a diamond seed, I suggest that six tests are needed. Their order matters because each test addresses a specific possible error and depends on the preceding tests.

The first is to calibrate the reservoir coordinate. A graphite coupon at equilibrium fixes $S_G = 1$, and that single point is not enough, because every claim made here concerns activities above it. At least one independent route to $S_G > 1$ must therefore be validated for the chosen flux; the candidate routes are set out below. Until one of them is in place, “supersaturated” describes the gas feed and not the liquid.

The second test is to bracket the seed front from both sides. Find the condition at which a registered diamond front stops retreating and begins to advance, and approach that condition from the carbon-poor and the carbon-rich side in the same cell. Arriving at the same value from both directions is what separates a thermodynamic boundary from a rate that has fallen below the detection limit. A one-sided experiment cannot make that separation, however carefully the front is measured.

The third test is to show both that the carbon came from the reservoir and that the interface moved. Detection of the source isotope label in confirmed diamond shows that carbon from that source entered the diamond. Crystallographic continuity establishes that the new material is part of the seed. A fiducial outside the reactive region establishes that the interface now lies beyond where it started. All three are required, because any two of them are also consistent with a separate crystallite that nucleated and grew against the seed.

The fourth test is to measure the reversible carbon-transfer boundary of one finite graphite competitor whose layer count, edge structure, support, strain, and passivation are all specified. This is the only comparison in this Perspective that can produce an operating interval, and it is pairwise: it bounds that one object and nothing else.

The fifth test is to quantify competitors that appear at random. Report the area watched, the time watched, and the efficiency with which a graphite particle of the relevant size would have been detected and recovered. Given those three quantities, a run in which no graphite was found bounds an event rate. Without them it carries no quantitative information.

The sixth test is to close the phase-resolved carbon balance before combining results or attempting scale-up. Diamond, graphite, amorphous carbon, carbides, dissolved carbon, wall deposits, and losses to the gas should sum to the carbon fed, within a stated uncertainty. A program that passes the first five tests but cannot close this inventory has not established where the carbon goes. Until the carbon balance closes in the present geometry, scale-up would introduce additional loss paths and make the result more difficult to interpret.

A failed test identifies an inadequate assumption or measurement. If the two-sided bracket of the second test does not converge, the front is under kinetic rather than thermodynamic control at those conditions. If the fourth test returns a finite-graphite boundary at or below the diamond repeat threshold, the pairwise interval of Section 3 has no solution for that competitor, and that route to an operating interval is closed. SI Section S4 specifies six corresponding protocols with their controls, false positives, success criteria, and falsifying outcomes.

The finite-competitor test concerns one specified graphite object. A graphite “witness” is a pre-existing, characterized graphite object deliberately exposed to the chosen flux, preferably in a matched cell separate from the diamond seed. It should be characterized before and after exposure, and its reversible boundary should be approached from both growth and dissolution sides. The most relevant witnesses reproduce the graphite states formed by the system itself; deliberately modified witnesses can test whether a proposed atom-scale contribution is large enough to matter. The response of one witness does not show that other graphite states, amorphous carbon, carbides, or off-seed nuclei are suppressed. Those competitors require their own rate and inventory measurements. Figure 5 summarizes the proposed experiments and distinguishes the reversible behavior of a prepared graphite state from the random appearance of competing carbon phases.

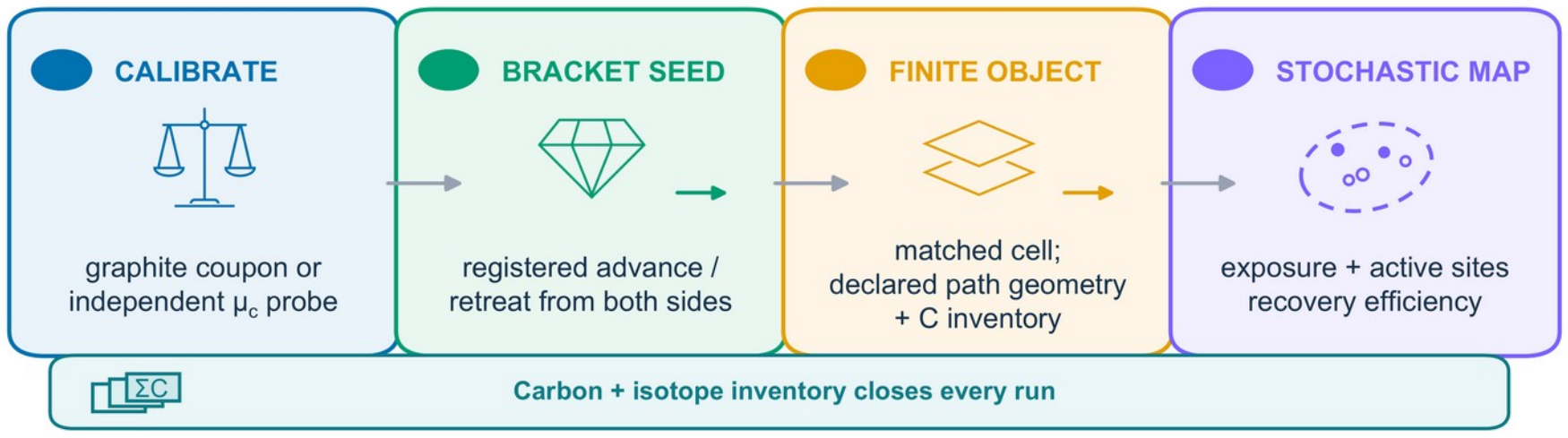

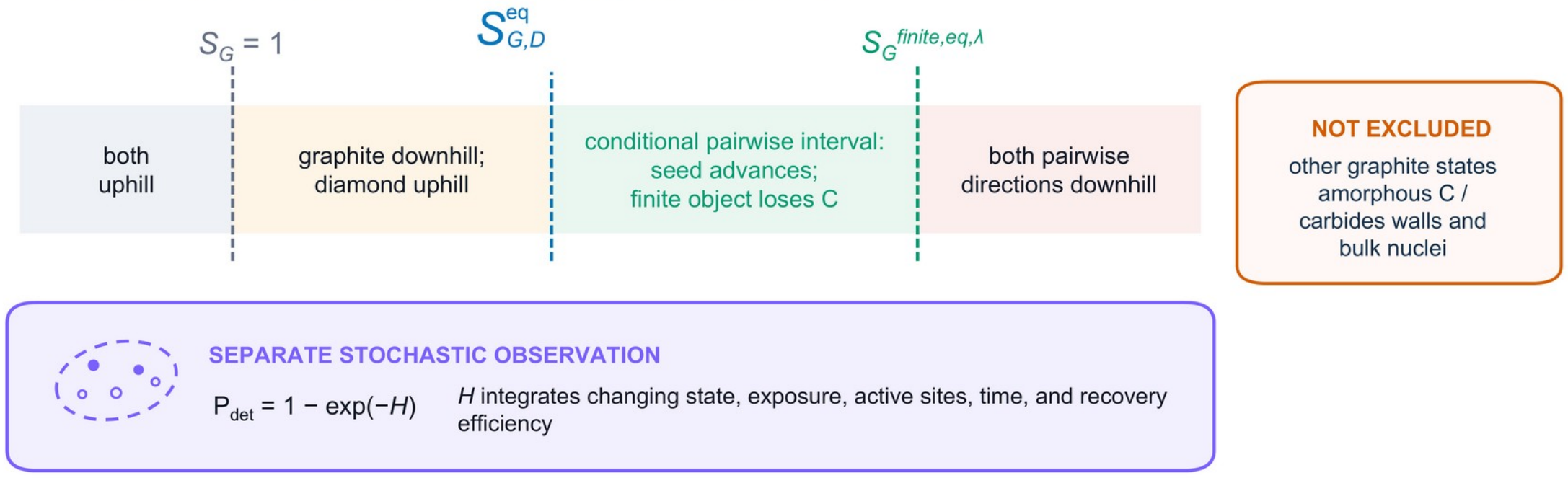

Figure 5. General test sequence and distinct stochastic observation. (a) For a chosen flux M, graphite calibration, the supersaturated diamond-seed scan, the controlled finite-graphite test, and carbon-inventory closure require separate experiments or independently controlled measurements. (b) The conditional pairwise boundary of one declared finite graphite path is distinct from the integrated stochastic nucleation hazard; nondetection bounds a protocol-specific rate rather than locating equilibrium. In panel b, α denotes the specified competitor pathway introduced above. Hα is its integrated detected-event hazard, and Prα is the probability of detecting at least one event through that pathway, as defined in SI Equation S22. The figure labels these quantities H and Pdet, respectively.

Above graphite saturation, the numerical $S_G$ scale requires more than the single equilibrium point. Candidate routes include a thermodynamic activity model fitted to independent equilibrium data, a validated gas-liquid equilibrium, or an electrochemical carbon-potential probe. Until one route is validated for the chosen $M$, $S_G$ above one is model-dependent. The clearest pairwise result would be registered diamond advance at the same local $S_G$ at which a declared finite graphite object reversibly loses carbon. Even that result would not establish global carbon-phase selection; other carbon states and spontaneous nuclei remain separate competitors.

Calibration should report uncertainty in temperature, composition, gas conversion, equilibrium assumptions, and the activity model. It should also test whether the same carbon-potential value is obtained from independent routes. Agreement only in nominal feed composition is insufficient because different fluxes, or the same flux after composition drift, can assign different carbon chemical potentials to that composition.

## 6. BN requires additional coordinates

BN cannot be described by one elemental chemical potential. Choose explicit elemental references and separate the BN-pair coordinate from B-rich versus N-rich imbalance:

$$\Delta\mu_B=\mu_B-\mu_B^{\mathrm{ref}}, \qquad \Delta\mu_N=\mu_N-\mu_N^{\mathrm{ref}}.$$

$$\Delta\mu_{\mathrm{pair}}=\Delta\mu_B+\Delta\mu_N, \qquad \Delta\mu_{\mathrm{imb}}=\Delta\mu_B-\Delta\mu_N.$$

A neutral BN-pair event changes the pair amount by one and the imbalance by zero. A 1:1 feed fixes the delivered atomic ratio, not the interfacial imbalance, terminations, antisites, vacancies, borides, or nitrides. When a declared event exchanges electrons with a reservoir, the electron electrochemical potential and the electrical boundary condition must also be specified; charged-supercell calculations require finite-size and potential-alignment corrections.[41]

These coordinates keep track of the reservoirs; they do not tell us which BN phase will grow. Two-dimensional hBN calculations show how elemental imbalance can change edge advance, but that supported monomer-growth problem does not establish a three-dimensional cBN repeat or a cBN-over-layered-BN interval.[42] High-level calculations also disagree on cBN/hBN ordering.[43,44] Metal-solvent cBN synthesis, molten-metal hBN growth, in situ hBN kinetics, and hydrogen-dependent BN nucleation calculations identify relevant conditions and barriers; they do not establish low-pressure cBN selection.[45–48] A credible test requires pair and imbalance control, declared electrical conditions, explicit finite-state comparisons, and rate-resolved competition. SI Section S5 gives the complete construction.

Compatibility does not establish selection. Selection asks which motif grows. Chemical-potential inequalities can identify conditions consistent with stoichiometric pair incorporation and selected defect populations. They do not compare all relevant finite cBN, wBN, and layered-BN states, nor do they supply the barriers and exposures that determine which motif appears. Mapping those conditions therefore bounds where cBN growth is allowed. Which motif appears within that region is a separate measurement. Figure 6 summarizes the BN compatibility coordinates and the additional requirements for motif selection.

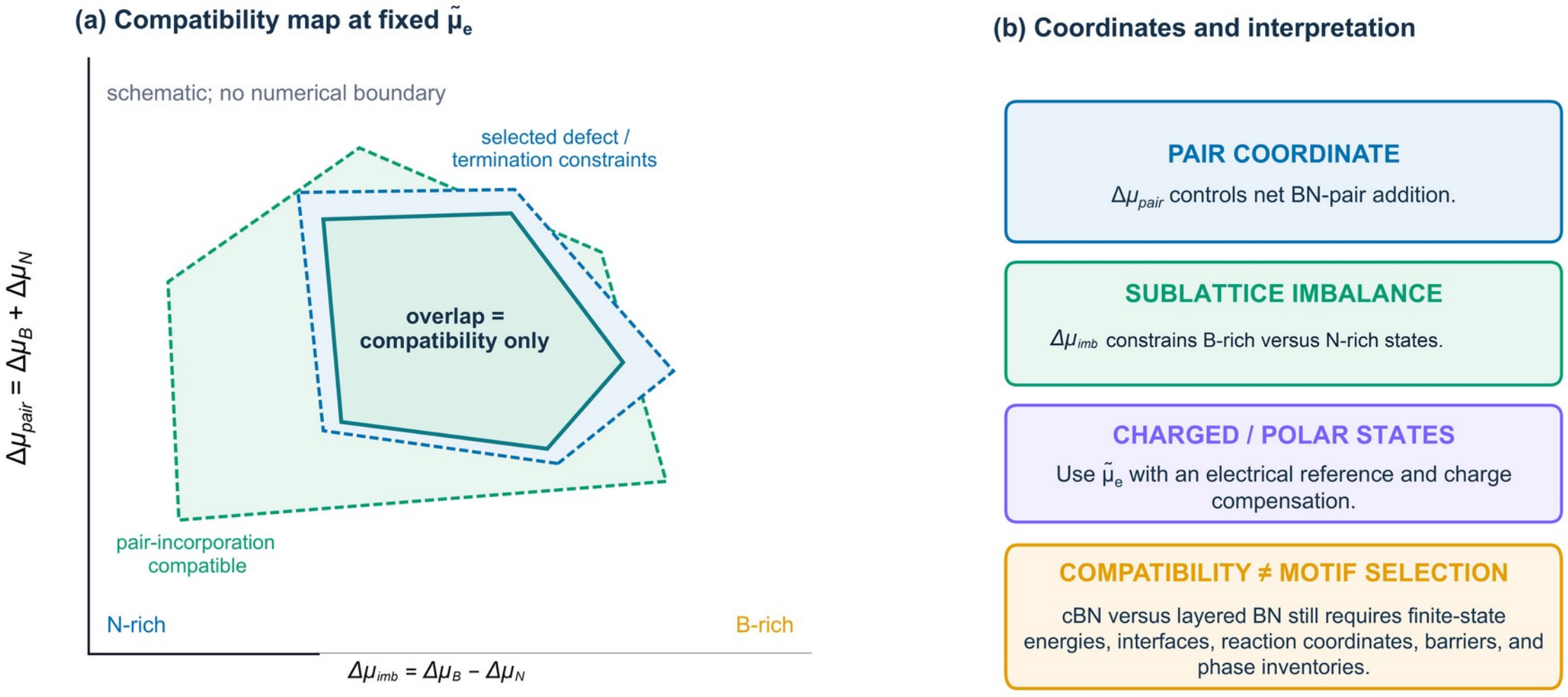

**Figure 6. BN compatibility coordinates do not select the observed motif.** (a) A schematic slice separates BN-pair chemical potential, B/N sublattice imbalance, and the electron electrochemical potential for states that exchange electrons with a reservoir. The panel-b label "charged / polar states" is broader than the coordinate's validity: $\tilde{\mu}_e$ enters the bookkeeping only when the declared event exchanges electrons with a reservoir, and other polar-interface ensembles require their own electrical boundary conditions and work terms. Dashed regions denote only compatibility constraints; no numerical boundary is implied. (b) Pair/imbalance compatibility, charged-state energetics, finite-state/interface energies, and barriers/rates are successive requirements. Their intersection does not by itself establish cBN over layered BN.

## 7. Thin metal films separate two kinetic boundaries

A thin film of a chosen carbon-bearing metal or alloy on diamond, supplied from the gas side, shortens the transport distance and provides a large buried solid-liquid area. It nevertheless contains two distinct boundaries—gas/metal and metal/diamond—separated by a finite, nonideal, possibly multicomponent reaction-diffusion medium. A short diffusion time does not show that diffusion controls the rate, that chemical potential is uniform, or that either boundary is at equilibrium.

Gas uptake and desorption, homogeneous reactions, multicomponent diffusion, thermodiffusion, buried-interface attachment and detachment, dewetting, passivation, and competing phases can each control the response. A fitted overall rate constant cannot distinguish these processes. Several independent perturbations and a closed carbon balance are required.

In the restricted passive, isothermal, diagonal-response limit, carbon flux feeding a buried macroscopic diamond front implies a supplying free-surface carbon potential at least as high as that at the buried interface. Because the diamond repeat requires $S_G > S_{G,D}^{eq} > 1$, suppression of graphite or amorphous carbon at that free surface is then kinetic. In the general multicomponent problem, Onsager cross-coefficients and thermal forces can drive carbon against its own chemical-potential gradient, so the coupled transport equations must be solved before inferring the profile.[49,50]

SiC vapor-liquid-solid growth and Si or Ge liquid-phase epitaxy demonstrate condensed-reservoir geometry, seeded growth, and useful interface measurements, but not selection between $sp^2$ and $sp^3$ carbon networks.[51–56] Seeded Na-flux GaN gives a closer process analogy: experiments separately resolve seed loss, seed-only growth, off-seed nucleation, polarity, gas-liquid supply, residual-film history, step bunching, inclusions, impurities, and recovered inventories.[57–62] These precedents identify measurements worth making. They do not supply a carbon activity or a growth boundary that can be carried over to diamond. A thin-film diamond experiment must independently validate wetting, film continuity, composition drift, gas-side chemistry, free-surface carbon formation, direct gas access to the seed, buried-interface motion, and complete inventories. SI Section S6 gives the conservation equations and precedent-specific limits.

The GaN literature also shows why finite-time process maps cannot be read as equilibrium phase diagrams. Reported boundaries changed with seed polarity, gas-liquid area, off-seed nucleation, melt history, and the method used to measure solubility. A 2026 comparison reduced off-seed polycrystalline GaN from about 11% to about 1% of the source gallium while temperature and pressure changed together; supersaturation was inferred rather than calibrated.[62] That process result is useful. No transferable activity threshold can be taken from it. For diamond, seed motion, off-seed competitors, wetting, gas-side products, and phase-resolved inventories must be scored separately.

## 8. Conclusions

The analysis applies to a viable pre-existing diamond seed in contact with a carbon-bearing metal flux when the reservoirs, external controls, front state, carbon-delivery chemistry, and relevant competitors can be defined. $\Delta Y$ accounts exactly for the specified transfers within that declared ensemble. Its sign gives thermodynamic direction for the stated event, and its completed-repeat limit recovers the bulk free energy per growth unit while retaining nonrepeat contributions exactly once.

There is no ambiguity in the macroscopic constraint. Bulk graphite saturation occurs at $S_G = 1$, whereas the pure macroscopic diamond threshold is $S_{G,D}^{eq}$, about 2.0 at 1300 K with the thermochemical data and model adopted here. The chosen flux changes the mapping from composition and feed onto that scale, and it changes finite states, barriers, transport, and losses; it does not create a macroscopic diamond-only thermodynamic interval.

A pairwise finite-competitor interval would require registered diamond advance at the same local $S_G$ at which one fully specified finite graphite state reversibly loses carbon, together with isotope-attributed outward motion and common-basis inventory closure. For BN, pair, imbalance, and applicable electronic coordinates must be controlled without mistaking defect compatibility for motif selection.

ΔY is an accounting identity. The experiments test whether the chosen states, chemical-potential model, finite-object boundary, barriers, transport model, and product inventories describe the actual system. If the independently measured boundaries, rates, isotope profiles, and carbon inventories cannot be reconciled within their uncertainties, at least one of those assignments is wrong.

## Author Information

**Corresponding Author.** Rodney S. Ruoff—Center for Multidimensional Carbon Materials (CMCM), Institute for Basic Science (IBS), Ulsan 44919, Republic of Korea; Department of Chemistry, Department of Materials Science and Engineering, and School of Energy and Chemical Engineering, UNIST, Ulsan 44919, Republic of Korea. Email: ruofflab@gmail.com.

**Biography.** Rodney S. Ruoff is a UNIST Distinguished Professor and founding Director of the IBS Center for Multidimensional Carbon Materials. His research concerns the chemistry and physics of carbon and related materials, including graphene, carbon nanotubes, porous carbons, diamanes, diamond, hBN, and cBN, with recent emphasis on studies of carbon, boron, and/or nitrogen dissolved in liquid metals and on low-pressure metal-flux growth. He received his Ph.D. in chemical physics from the University of Illinois at Urbana-Champaign and held positions at SRI International, Washington University, Northwestern University, and the University of Texas at Austin before joining UNIST.

## Supporting Information

Definitions and ensemble controls; carbon thermochemistry and reservoir-transformed finite-state derivations; kinetic and isotope interpretation; common-basis inventory accounting; six experimental protocols with hypotheses, false positives, controls, success criteria, and falsifiers; BN coordinates and charge constraints; a finite-film reaction-diffusion model; and a symbol glossary with reporting checklist (PDF).

## Notes

Competing financial interests. The Institute for Basic Science filed Korean Patent Application 10-2023-0052752, which includes Rodney S. Ruoff as an inventor. A second patent application based on related concepts, “Transport of carbon, nitrogen and boron,” Korean Patent Application 10-2023-0184624 (filed December 18, 2023), includes R. S. Ruoff as an inventor. The author declares no other competing financial interest.

## Acknowledgments

This work was supported by the Institute for Basic Science (IBS-R019-D1). Hirotomo Nishihara, James E. Butler, Da Luo, Yanwu Zhu, Kun Li, and Wenhao Sun read and commented on earlier drafts of the manuscript. The author thanks Daniel Hedman and Pavel Bakharev for discussions and Daniel Ogenrwot, Md Mahbubul Alam, Adnan Yousaf, Sudipta Bag, and Vijay Gupta for providing requested articles, critically reading the manuscript, and discussing the group’s metal-flux research.

The author also thanks Alisher Sultangaziyev for preparing and revising all manuscript figures with assistance from Claude (Anthropic) for graphical design and refinement. The scientific content of the figures was specified, reviewed, and approved by the author.

During preparation and revision of this manuscript, the author used Anthropic Claude and OpenAI ChatGPT/Codex to assist with literature organization; drafting and revision of text; development and checking of derivations, equations, calculations, and programmatic figure drafts; and checks of structure, notation, internal consistency, and references. The author directed the scientific scope and all revisions, reviewed and revised all AI-assisted material, and takes full responsibility for the final text, equations, calculations, citations, and graphics.

# Supporting Information

## Thermodynamic and Kinetic Tests for Low-Pressure Metal-Flux Growth of Diamond, with an Extension to Cubic Boron Nitride

**Rodney S. Ruoff**

Center for Multidimensional Carbon Materials (CMCM), Institute for Basic Science (IBS), Ulsan 44919, Republic of Korea; Department of Chemistry, Department of Materials Science and Engineering, and School of Energy and Chemical Engineering, UNIST, Ulsan 44919, Republic of Korea.

## Overview and how to use this document

The main Perspective explains the physical argument. This Supporting Information supplies the formal definitions, derivations, boundary conditions, measurement logic, and experimental protocols needed to apply or test that argument.

The shortened main text retains the central event definition, macroscopic threshold, pairwise finite-state test, evidentiary ladder, and experimental sequence. Historical development, full derivations, rate and inventory equations, protocol controls, BN coordinates, and thin-film transport and precedent analyses are consolidated here in Sections S1–S6. This organization reduces repetition without changing the scientific scope of the source pair.

The analysis considers a pre-existing diamond seed in contact with a **chosen carbon-bearing metal flux**, denoted *M*. Here, the flux is a metal or alloy that is liquid under the growth conditions. *M* denotes only the metal or alloy that constitutes the flux; its local composition and reservoir/interfacial state relevant to the event are separate quantities that must be specified independently. The accounting is not restricted to Ga-Fe-Ni-Si, but it applies only when the seed remains viable, carbon can reach the interface, the independent reservoir components and external controls can be declared, relevant competing phases and reactions can be identified, and the interface can be interrogated reproducibly enough to test growth, retreat, and inventory.

The 2024 Ga-Fe-Ni-Si system under methane/hydrogen feeding is the demonstrated reservoir system.[1] It supplies the experimental basis and system-specific constraints, but not a universal activity model, kinetic mechanism, or apparatus for another flux. Every new *M* requires its own carbon activity-composition relation, dissolved-species chemistry and equilibration behavior, metal-diamond interfacial states, competitors, barriers, transport coefficients, wetting and film stability, gas-liquid chemistry, and loss or passivation pathways.

A single example runs through the document. A diamond seed is in contact with *M*, which contains or receives carbon. Carbon may attach to the diamond front, detach from it, remain dissolved, form graphite, carbide, or amorphous carbon, or leave through another pathway. The analysis specifies the constrained front transition and distinguishes its thermodynamic direction from its reaction-coordinate-dependent barrier, observed rate, and retained product inventory.

Unless a section is explicitly labeled **per particle**, chemical potentials, free energies, and activation free energies are molar quantities in J mol$^{-1}$ of the declared growth unit or event basis. The carbon growth unit is one mole of C atoms. The default BN growth unit is one mole of neutral BN pairs. Molar area and volume per mole of growth units are written with overbars, for example $\bar{A}_u$ and $\bar{V}_u$, and the thermal factor is $RT$. Atomistic equations in eV per atom or per BN pair are identified as such and use $k_B T$.

The equations include definitions and exact accounting identities, standard relations written for the declared variables, and constructions used for the present estimates. These categories are identified explicitly in Table S13 so that an identity is not mistaken for a new physical law and an illustrative construction is not mistaken for a measured material constant. Every important equation is followed by its physical meaning and limitations.

Section S1 defines one front event. Section S2 defines the carbon activity scale and distinguishes bulk from finite graphite. Section S3 connects event thermodynamics to rates, interface motion, inventories, and isotope evidence. Section S4 gives six general experimental protocols for a viable candidate *M*. Section S5 introduces the additional coordinates needed for BN. Section S6 treats a thin metal film as a two-boundary reaction-diffusion system. Section S7 gives a reporting checklist.

## S1. How to evaluate one event at an open growth front

### S1.1. Compare the same physical region before and after

**Purpose.** An event free energy is meaningful only if the initial and final states refer to the same physical control region under the same externally imposed controls and are states for which a reversible free energy is defined.

Choose a fixed interfacial control region at the diamond-$M$ interface that may contain part of a seed surface, steps, kinks, adsorbates, a finite nucleus, and a finite portion of the chosen metal flux. Its spatial boundaries and Gibbs dividing surfaces remain the same between the initial state $s$ and final state $s'$. The state description must include species inventories $N_i$, geometry, coverage, defects, strain, charge, polarity, and relevant interfacial phases.

In this SI, $s$ and $s'$ denote constrained-equilibrium states or reproducible metastable basins under declared temperature, pressure, mechanical, electrical, and reservoir controls. $G_{\mathrm{front}}(s)$ denotes the constrained reversible potential appropriate to that declared ensemble; the symbol $G$ is retained for readability, but any required mechanical or electrical Legendre transform is part of its definition. An arbitrary driven molecular snapshot does not automatically possess this free energy; genuinely nonequilibrium microstates require a stochastic-thermodynamic or other nonequilibrium construction and lie outside the exact-equilibrium claim made here.

Operationally, a reduced basin description is adequate only when the variables omitted from it relax on a timescale much shorter than the residence time in that basin. For a continuously advancing front treated quasi-statically, they must also relax before appreciable front motion. These tests determine whether the chosen state description is adequate; they make no assumption about every configuration along the transition path. A claimed reversible boundary should converge with hold time and scan rate and should be approached from both directions; any residual hysteresis must be reported.

At fixed declared controls, define the reservoir-transformed Gibbs state function

$$Y(s)=G_{\mathrm{front}}(s)-\sum_i N_i(s)\mu_i^{\mathrm{res}}. \qquad \text{(S1)}$$

The equation subtracts the chemical free energy of the material assigned to the reservoirs at the same inventories. The transformed quantity is useful because the control region is open: atoms can enter or leave while reservoir chemical potentials remain fixed.

Equations S1–S3 may be written in total quantities, with energies in joules and $N_i$ as absolute mole numbers. For the molar event normalization used here, every term is divided by the same amount of identical events. Then $Y$ and $\Delta Y$ are in J mol$^{-1}$ of declared events, and each $N_i$ or $\nu_i$ is in mol species $i$ per mol event.

Equation S1 is exact only for the declared ensemble and external controls. If a generalized force $X_j$ performs work through a change $\Delta\xi_j$, the transform must include the corresponding work term. Charged events require electrochemical potentials or explicit electrical work. Elastic events require the free energy appropriate to the stated stress or strain condition. Hidden work terms would otherwise be misidentified as chemical selection.

Changing a Gibbs dividing surface changes assigned interfacial inventories but must not change a correctly constructed event free energy. This invariance requires consistent bulk reference amounts, transfer stoichiometry, reservoir terms, and external-work terms in both states.

### S1.2. Subtract what the reservoirs supplied

**Purpose.** This subsection gives the thermodynamic direction criterion for one declared transition between constrained states.

Let

$$\nu_i=N_i(s')-N_i(s),$$

with $\nu_i>0$ for net transfer of species $i$ from its reservoir into the front. The finite-difference event quantity is

$$\Delta Y(s\longrightarrow s')=G_{\mathrm{front}}(s')-G_{\mathrm{front}}(s)-\sum_i \nu_i\mu_i^{\mathrm{res}}. \qquad \text{(S2)}$$

The first difference is the constrained reversible free-energy change of the control region. The sum subtracts the chemical free energy of everything supplied by the reservoirs. If one mole of C atoms enters the front, the carbon contribution is $-\mu_C^L$ per mole of events. If H, metal atoms, ions, or electrons also enter or leave, their stoichiometric electrochemical or chemical-potential terms must be included.

The affinity of the declared forward event is

$$\mathcal{A}\equiv-\Delta Y. \qquad \text{(S3)}$$

Therefore:

- $\Delta Y<0$: the specified forward event is thermodynamically downhill;
- $\Delta Y=0$: the event has zero affinity relative to its exact reverse;
- $\Delta Y>0$: the reverse event is thermodynamically favored.

This criterion gives thermodynamic direction and nothing more. It does not rank pathways unless each pathway uses consistent initial and final basins, event basis, reservoirs, and external controls.

Equations S1–S3 use a standard Gibbs-reservoir (semigrand) Legendre transform[2] and the forward-affinity sign convention stated in a modern thermodynamics source.[3] The contribution here is the explicit declaration of the crystal-front event, its completed-repeat bulk limit, the finite-competitor distinction, and the resulting measurement hierarchy.

For a matched Markovian elementary transition between the declared basins under local detailed balance, the conditional forward and reverse rate constants obey

$$(k_{s\to s'})/(k_{s'\to s})=\exp(-(\Delta Y)/(RT)). \qquad \text{(S4)}$$

An area-averaged one-way flux ratio has this form only when state populations, activities, and site occupancies are included consistently and no hidden parallel pathway or driven cycle is being coarse-grained away. At $\Delta Y=0$, matched forward and reverse one-way fluxes are equal for this reaction. Independent etching, transformation, passivation, transport loss, or nucleation elsewhere is not the reverse member of that pair.

Equation S4 is a **modern** local-detailed-balance statement for a matched elementary transition; it is not a formula attributed to the early crystal-growth papers. BCF applied the principle of detailed balancing to equilibrium kink configurations in a Kossel model, whereas Li *et al.* assessed modern attachment and detachment rate expressions against thermodynamic and detailed-balance criteria.[4,5] Molecular-scale experiments show that real attachment can contain several stages of desolvation, preliminary binding, and reorganization.[6]

## S1.3. Separate a completed lattice repeat from everything that remains

**Purpose.** The completed-repeat construction identifies the bulk-tracking part of a growth event and prevents finite terms from being counted twice.

Suppose one growth unit contains stoichiometric coefficients $b_i$ from the independent reservoirs, where $b_i$ is measured in moles of species $i$ per mole of growth units. Its reservoir chemical potential is

$$\mu_u^{\mathrm{res}}=\sum_i b_i\mu_i^{\mathrm{res}}. \qquad \text{(S5)}$$

For carbon, the simplest growth unit is one C atom, or one mole of C atoms on the molar basis. For BN, the default growth unit is one neutral BN pair. Other units, such as $C_2$ or a charged BN-containing complex, are allowed only when their stoichiometry and charge are stated.

A completed translational repeat begins and ends with congruent front basins separated by one lattice translation. “Completed” means that the entire growth unit has been incorporated; “translational” means that no new area, line, defect, reconstruction, adsorbate inventory, strain, charge, polarity, or configurational change remains after the repeat. The local kink has moved, but the final geometry is equivalent to the initial geometry.

This is a modern definition. The historical papers did not use $\Delta Y$, $\mu_{k,\varphi}^{\mathrm{rep}}$, or constrained-basin language. The repeat unit must also be complete: for a crystal whose growth unit has several subunits, attachment or detachment of all required subunits is needed before the crystal chemical potential is restored.

Define $\mu_{k,\varphi}^{\mathrm{rep}}$ as the reversible free-energy cost per mole of growth units for that strict event in phase $\varphi$. A general event involving $n$ growth units is partitioned as

$$\Delta Y = -n(\mu_u^{\mathrm{res}} - \mu_{k,\varphi}^{\mathrm{rep}}) + \Delta G_{\mathrm{nonrep}}. \qquad \text{(S6)}$$

The first term is the contribution of $n$ congruent repeats. The remainder $\Delta G_{\mathrm{nonrep}}$ contains only changes left over relative to those repeats: new area or edge length, curvature, reconstruction, retained adsorbates or metal atoms, defects, strain, charge, polarity, or configurational change. Each contribution appears once.

Any H, metal, ion, or electron exchanged while completing the repeat belongs in the transfer vector. If all non-product inventories return to their initial values and the product is pure, unstressed, and translationally homogeneous in the thermodynamic limit, with long-range step-step, support-field, and compositional-modulation excess per repeat vanishing, then

$$\mu_{k,\varphi}^{\mathrm{rep}} = g_\varphi^{\mathrm{bulk}}. \qquad \text{(S7)}$$

This equality does not apply unchanged to an alloyed product, stressed or curved front, surface compound, persistent support field, compositionally modulated front, or event that leaves a different adsorbate inventory.

For one fully labeled nonrepeat event, a compact effective cost can be defined as

$$\mu_{\mathrm{event}}^{\mathrm{eff}} = \mu_{k,\varphi}^{\mathrm{rep}} + (\Delta G_{\mathrm{nonrep}})/(n). \qquad \text{(S8)}$$

This event-specific quantity is not the strict repeat potential and cannot be used to imply that the metal has moved the pure macroscopic threshold.

**Table S1. Binding definitions and unit basis.**

| Quantity | Physical meaning | Unit and binding rule |
|---|---|---|
| $\Delta Y$ | Reservoir-transformed free-energy change between declared constrained front basins | J mol$^{-1}$ event basis; negative for the declared forward direction |
| $\mu_{k,\varphi}^{\mathrm{rep}}$ | Modern reservoir-transformed cost of a completed congruent repeat | J mol$^{-1}$ growth units; equals $g_\varphi^{\mathrm{bulk}}$ only in the stated homogeneous thermodynamic limit |
| $\Delta G_{\mathrm{nonrep}}$ | Remainder after subtracting congruent repeats | J mol$^{-1}$ event basis; each area, edge, curvature, defect, adsorbate, strain, charge, or geometry term appears once |
| $\beta$ | Edge or step line free energy | J m$^{-1}$; used for equilibrium edge excess |
| $\tilde{\beta}$ | Step stiffness, $\beta + \partial^2\beta/\partial\theta^2$; θ is the edge orientation | J m$^{-1}$; used in anisotropic-curvature and fluctuation formalisms |
| $\gamma$ | Interfacial free energy | J m$^{-2}$; specify interface, orientation, composition, stress, charge, and dividing-surface convention |
| $\Gamma_C$ | Carbon activity coefficient under the stated standard-state convention | Dimensionless |
| $\bar{A}_u$, $\bar{V}_u$ | Molar area and molar volume per mole of growth units | m$^2$ mol$^{-1}$ and m$^3$ mol$^{-1}$ |
| $A_{int}$ | Retained-flux interfacial area used for rate normalization | m$^2$ |
| $j$ | Molar interfacial or transport flux | mol growth units m$^{-2}$ s$^{-1}$ unless another basis is stated |
| $\mathcal{J}_\alpha$ | Activated-event or nucleation rate density | Events per stated exposure basis, such as m$^{-3}$ s$^{-1}$, m$^{-2}$ s$^{-1}$, or site$^{-1}$ s$^{-1}$ |
| $\mathcal{A}$ | Thermodynamic affinity of the declared forward event, defined as $-\Delta Y$ | J mol$^{-1}$ event basis; positive for the declared forward direction |
| $b_i$ | Stoichiometric coefficient of species i in one growth unit | mol species i per mol growth units |

| Quantity | Physical meaning | Unit and binding rule |
|---|---|---|
| $\Delta g_{DG}$ | Bulk diamond-minus-graphite molar Gibbs free-energy difference | J mol$^{-1}$ C |
| $S_G^{op}$ | Operating carbon activity on the graphite-normalized scale | Dimensionless |
| $H_\alpha$, $\mathrm{Pr}_\alpha$, $\mathcal{E}$, $\varepsilon_\alpha$ | Integrated detected-event hazard, probability of at least one detected event, active exposure, and detection/recovery efficiency | $H_\alpha$, $\mathrm{Pr}_\alpha$, and $\varepsilon_\alpha$ are dimensionless; $\mathcal{E}$ has the declared exposure basis |
| $A$, $\mathcal{P}$, $\ell$; $A_C$, $\bar{A}_C$ | Feature area, perimeter, A/$\mathcal{P}$, basal area per C atom, and its molar value | m$^2$; m; m; m$^2$ per C atom; m$^2$ mol$^{-1}$ C |

## S1.4. Historical lineage and the limit of equilibrium kink statistics

**Purpose.** This subsection distinguishes what the historical sources established from the reservoir-transformed analysis used here.

Kossel analyzed the energies of successive lattice-building operations and observed that an attachment process whose repetition constructs the complete crystal must deliver the same energy per building unit. Stranski singled out a surface position whose attachment and removal can be repeated indefinitely and argued that its detachment work determines saturation. The site became known as the **half-crystal position**; related constructions appear in Kossel and Stranski. Kaischew analyzed the **repeatable growth step** and set its forward and reverse probabilities equal at equilibrium. BCF then developed the terrace-step-kink description quantitatively: kinks are step positions from which the evaporation energy equals the crystal evaporation energy; their equilibrium statistics control step structure, while supersaturation, diffusion, two-dimensional nucleation, and dislocations control step motion and sustained growth.[4,7–9]

The present $\Delta Y$ construction, constrained front basins, and $\mu_{k,\varphi}^{rep}$ are modern formalizations consistent with that lineage. In particular, the early papers did not use the modern kink and reservoir-transformed notation, and BCF did not formulate the present event quantity.

BCF separated equilibrium step structure from growth kinetics and noted that finite kink spacing or slow exchange can limit step motion. De Yoreo et al. demonstrated a weak-fluctuation regime in calcite in which sparse kinks were replenished too slowly during growth; kink-pair creation, propagation, and impurity blocking then controlled the rate.[4,10] This limits quasi-equilibrium kink statistics in kinetic closures. The thermodynamic repeat-site relation for a declared event is unaffected. Ehrlich and Hudda established a terrace-edge barrier.[11] In the Schwoebel-Shipsey model and sign convention, conventional easier lower-terrace attachment stabilizes step spacing during growth; the inverse asymmetry is needed for growth bunching.[12] Their conventional asymmetry produces bunching on the evaporation branch, whereas the Bales-Zangwill analysis predicts a meandering instability during net deposition.[13] The published erratum to the latter is cited with the original. A kink Ehrlich-Schwoebel effect can instead control in-plane step morphology.[14] These effects are morphological and kinetic in origin. They say nothing about equilibrium kink densities. Decay of metastable one-dimensional structures on Si(111) scaled with an exponent of 4.3 ± 0.5 and was interpreted as locally conserved mass transport.[15] Jeong and Williams review the step stiffness used below.[16]

Table S2. Historical sources and their role in the present analysis.

| Source | Direct contribution | Regime and assumptions | Inherited concept | Modern or later element not attributed to the source |
|---|---|---|---|---|
| Kossel, 1927 | Energetic hierarchy of lattice-building operations; repeated operations construct the crystal | Idealized ionic lattice; molecular energetics; explicitly questions an overly coarse equilibrium-surface-energy treatment | Repetition as the route from local attachment to the whole lattice | Constrained basins, $\Delta Y$, and $\mu_{k,\varphi}^{rep}$ |
| Stranski, 1928 | Special surface position whose attachment/removal | Near equilibrium; own dilute vapor; simplified surface | Repeatable equilibrium position and bulk-tracking | Modern kink notation, reservoir transforms, and arbitrary driven- |

| Source | Direct contribution | Regime and assumptions | Inherited concept | Modern or later element not attributed to the source |
|---|---|---|---|---|
| | can repeat indefinitely; its detachment work fixes saturation | structure and neglected thermal motion in detachment-work estimates | detachment work | interface kinetics |
| Kaischew, 1936 | Kinetic analysis of the half-crystal position and repeatable growth step; equal attachment and detachment probabilities at equilibrium | Gas concentration not too far from equilibrium; kinetic probability treatment | Kinetic equal-probability treatment of the half-crystal position and repeatable step | Local detailed balance for coarse-grained driven networks and modern state-function notation |
| BCF, 1951 | Equilibrium kink statistics; step motion; diffusion; two-dimensional nucleation; dislocation-supplied steps; detailed balancing in kink statistics | Kossel working model plus stated extensions; equilibrium step structure used in kinetic growth theory; acknowledges kink-spacing and exchange limits | Quantitative terrace-step-kink and step-flow framework | Present completed-repeat potential and claim that every real interface maintains quasi-equilibrium kink statistics |
| De Yoreo *et al.*, 2009 | Experimental and kMC demonstration of kink-limited growth and impurity blocking in weakly fluctuating steps | Calcite and related low-solubility/low-temperature systems; finite observation times; sparse kinks | Validity limit for rapid-fluctuation/high-kink-density closures | Rejection of all BCF theory or of the thermodynamic repeat-site identity |
| van der Putte, van Enckevort, Giling and Bloem 1978; Bloem and Giling 1978 (silicon CVD and HCl etching) | An analytic smooth-to-bunched process boundary in driving force against reciprocal temperature whose position is set by the kink formation energy through the Burton–Cabrera mean kink spacing, compared with a measured activation energy of 87 kcal mol$^{-1}$ (the review instead used 85 kcal mol$^{-1}$ with a different energy partition). Step density treated as a finite resource whose exhaustion forces two-dimensional nucleation. | Vapor growth and HCl etching of silicon at one atmosphere; a single polymorph; the kink formation energy assumed equal to the Si–Si bond energy rather than computed; the quantitative calibration taken from etching, where the sign of the controlling inequality is reversed relative to growth. | The form of the criterion—a kink statistic multiplying a transport comparison—and the reading of a morphological transition as a kinetic process boundary rather than a phase boundary. | Computed kink formation energies, selection between two competing solid polymorphs, and reservoir-transformed treatment of a condensed carbon-bearing flux. |
| Bales and Zangwill 1990 | Linear stability analysis showing that a terrace edge advancing under net deposition meanders | Growth only, with a deposition flux strictly above its equilibrium value; a single component; an | The distinction between an instability of the spacing between steps and an instability of the in- | Any statement about kink density; the result is a Mullins–Sekerka analogue and is silent about kink |

| Source | Direct contribution | Regime and assumptions | Inherited concept | Modern or later element not attributed to the source |
|---|---|---|---|---|
| | when attachment is easier from the lower terrace. | isolated step or a train whose diffusion fields are treated as independent. | plane shape of one step. | supply. |
| Vekilov 2007[17] | Across about ten chemically diverse solution-grown systems, an enthalpic barrier contribution of 28 ± 7 kJ $mol^{-1}$ compiled by that review from a conference report; near 298 K the stated range corresponds to roughly four to six orders of magnitude relative to a barrierless reference | Aqueous and organic solution growth of salts, organic crystals, proteins, and viruses; the slowdown is a derived near-298 K estimate | That a condensed reservoir imposes an attachment barrier absent from vapor growth, so the kinetic coefficient is a property of the reservoir as well as of the surface. | Any value for a metal flux. No metallic solvent appears in the compilation; carrying the magnitude over to Ni, Fe, or Cu is introduced here and remains untested. |
| Ju et al. 2022[18] | Analytic BCF generalization for alternating inequivalent step types, relating terrace-side attachment asymmetry to line diffusion and kink attachment | Vapor-phase growth; analytic and continuum throughout; a single component; no polymorph competition. | That inequivalent steps on one surface require separate kinetic coefficients, and that the Ehrlich–Schwoebel asymmetry may reside at the kink. | The binary chemical-potential coordinates and the polymorph-selection criterion. The treatment is analytic, contains no kink formation energies, and involves no machine learning. |
| Zhang et al. 2016[19] | The binary growth problem written in supersaturation and composition coordinates for two-dimensional hBN. The anisotropy of one-dimensional row nucleation changes by about 1.9 eV across a 1 eV window in $\mu_B$ at essentially fixed $\Delta\mu_{BN}$. In that source, $\mu_B + \mu_N = \mu_{BN}^0 + \Delta\mu_{BN}$, with $\mu_{BN}^0$ the energy of one BN pair in an hBN sheet on the metal. | Two-dimensional hBN on a catalytic metal substrate; monomer incorporation; no cubic phase and no three-dimensional growth front. | The coordinate split: a binary system has one supersaturation-like and one composition-like degree of freedom, and the element held in excess selects which edge advances. The source-specific $\Delta\mu_{BN}$ is not identical to the present $\Delta\mu_{pair}$ unless the reference states are matched. | Extension to a three-dimensional cubic motif, the requirement that the repeat unit be a B–N pair, and the diamond-versus-graphite comparison. |

## S1.5. A 1978 precedent: a kink-statistical process boundary for silicon

The construction used in the main text—a kink statistic fixing a process boundary in a plane of driving force against reciprocal temperature—has a close structural precedent in the silicon literature. Bloem and Giling, together with van der Putte and co-workers, derived a kinetic boundary for the smooth-to-bunched transition of an HCl-etched silicon surface.[20–22] I describe it because the correspondence is useful and the differences define what is new here.

First, the terrace problem is of BCF type but uses a first-order kinetic boundary condition at the step rather than a perfect sink. With surface diffusion length $x_s = (D_s\tau)^{1/2}$, the flux collected across a terrace of half-width $L$ carries $\tanh(L/x_s)$. In relabeled local notation used here, $a_{lat}$ is a lattice length, $\sigma_{rel}$ is relative supersaturation, and $\mathcal{J}_N$ is incident particle-number flux; the vertical rate is $v_z = 2a_{lat}^3\sigma_{rel}\mathcal{J}_N x_s \tanh(L/x_s)/L$. The far-step and close-step limits follow directly. These local symbols are not the interfacial free energy $\gamma$ or molar flux $j$ used elsewhere in this SI.

Second, the source writes $x_s = a_{\mathrm{lat}} \exp[(E_{\mathrm{des}} - E_{\mathrm{diff}})/(2k_BT)]$ for per-particle energies. With $E_{\mathrm{des}} > E_{\mathrm{diff}}$, $x_s$ decreases as temperature rises. In that model, a widely spaced step train can therefore show a rate that decreases with temperature; the behavior was interpreted in that system as step-limited incorporation rather than bulk transport. Molar energies quoted below use $RT$, not $k_BT$.

Third, the prediction was compared with experiment. In vacuum silane pyrolysis, the incorporated fraction was interpreted to have an apparent negative activation energy with a magnitude of 7 ± 2.5 kcal mol$^{-1}$. Bloem and Giling's review identified this magnitude with $(E_{des} - E_{diff})/2$ and inferred $E_{des} \approx 50$ and $E_{diff} \approx 36$ kcal mol$^{-1}$. They interpreted the negative branch as failure of arriving atoms to reach a stable kink before desorption.

Fourth, the driving force at the step is not the nominal bulk value. With interface rate coefficient $k_d$, boundary-layer thickness $\delta$, and diffusivity $D$, define the dimensionless transport number $N_{tr} \equiv k_d\delta/D$. The surface supersaturation is the bulk value divided by $1 + N_{tr}$. The authors also showed that replacing $(c_b - c_s)$ by $(c_b - c_{eq})$ in the transport-flux expression incurs about 10% error at $N_{tr} = 10$. This is distinct from replacing surface supersaturation by bulk supersaturation.

Fifth, the model assumes that bunching begins when the imposed step velocity approaches the adatom diffusion velocity and then weights that condition by the mean kink spacing. On a molar-energy basis, the reported transition line is $p_{\mathrm{SiCl2}}^{(\mathrm{eq})} = C \exp[-(E_{\mathrm{diff}} + E_b)/(RT)]$. The reconstructed Arrhenius line gave $E_{\mathrm{diff}} + E_b = 87$ kcal mol$^{-1}$. Assuming $E_b = 55$ kcal mol$^{-1}$ as the Si–Si bond energy leaves 32 kcal mol$^{-1}$ for surface diffusion. The structure, not a unique partition of the total, is the useful precedent: a kink statistic fixes where the boundary lies, and a transport comparison selects its branch.

Sixth, step density is treated as a finite step resource and, by analogy, kink supply can be exhausted. Where the imposed rate exceeds what the existing steps can carry, the deficit must be made up by two-dimensional nucleation, and this was observed: on (111) surfaces misoriented by less than about half a degree, beyond a critical HCl concentration the surface broke up into shallow triangular facets inclined by about one degree from the true (111), because, in their words, the original surface possesses too few steps to respond properly to the desired etch rate. The surface, as they put it, creates its own misorientation. That is the silicon precedent for treating a kink population as a resource that can be exhausted rather than as a boundary condition that always accommodates the imposed rate.

One numerical caveat has to be recorded before this precedent is used, because it bears on how much weight the agreement can carry: the companion review restates the same result differently.

There the transition line is quoted as 85 rather than 87 kcal mol$^{-1}$, and it is compared not with $E_{diff} + E_b$ but with $\tfrac{1}{2}(2E_b + E_{diff} + E_{des}) = \tfrac{1}{2}(110 + 15 + 45)$ kcal mol$^{-1}$, that is with $E_b \approx 55$, $E_{diff} \approx 15$ and $E_{des} \approx 45$ kcal mol$^{-1}$. A desorption energy enters the review's partition that cancels out of the primary derivation, and the diffusion energy it uses, 15 kcal mol$^{-1}$, is neither the 32 kcal mol$^{-1}$ the primary analysis returns nor the roughly 36 kcal mol$^{-1}$ implied by the review's own low-temperature growth calibration. The review's 45/15 partition also changes $E_{des}$ from about 50 to about 45 kcal mol$^{-1}$ and gives $(E_{des} - E_{diff})/2 = 15$ kcal mol$^{-1}$, not the reported 7 ± 2.5 kcal mol$^{-1}$ magnitude. The two routes give similar totals because both assign about 55 kcal mol$^{-1}$ to the kink and about 30 kcal mol$^{-1}$ to transport.

I therefore use the primary paper as the source of the number and interpret the agreement as structural.

Five limits bound how far this precedent can be carried.

First, the treatment contains no computed kink formation energy. $E_b$ was identified with the silicon–silicon bond energy rather than calculated, so the kink term is an input to the decomposition and not a test of it, and the one quantity actually extracted from the data, $E_{\mathrm{diff}} = 32$ kcal mol$^{-1}$, was checked only against a literature range spanning 5 to 35.

Second, there is no competing polymorph and no reservoir transformation, so nothing in it addresses selection between two solid phases.

Third, the energies entering the decomposition are assumed values, and as noted above they differ between the primary paper and the review, so the numerical agreement should be read as structural rather than quantitative.

Fourth, the transition-line data come from HCl etching and not from growth, and the direction of the controlling inequality reverses between the two.

Fifth, in BCF's notation, where $a_{step}$ is the intermolecular distance along the step and $w$ is the energy required to form one kink, the mean kink spacing is $x_0 = (a_{step}/2)[\exp(w/k_BT) + 2] \approx (a_{step}/2)\exp(w/k_BT)$ for $w/k_BT \gg 1$. The exponent is not $w/(2k_BT)$. The criterion above and the one in the main text use the large-spacing form.

# S2. Carbon reference scale and the distinction between bulk and finite graphite

## S2.1. Use bulk graphite saturation as the experimental reference

**Purpose.** This subsection defines a dimensionless carbon-potential scale and states when one elemental carbon chemical potential is a valid description.

For carbon in the standard state declared for the chosen metal-flux reservoir $M$,

$$\mu_C^L = \mu_C^{\circ,L}(T,P,M) + RT\ln a_C. \qquad \text{(S9)}$$

The activity $a_C$ is dimensionless. In a nonideal alloy, $a_C = \Gamma_C x_C$ only after the standard state and activity-coefficient convention have been stated. Carbon mole fraction alone does not determine chemical potential.

A single elemental $\mu_C^L$ is appropriate when the dissolved carbon-bearing species interconvert rapidly enough to share local chemical equilibrium on the transport and attachment timescales. If methane-derived fragments, carbides, clusters, or other carbon species do not equilibrate locally, the relevant description requires species-specific chemical potentials and reaction affinities rather than one scalar $\mu_C^L$.

At equilibrium with macroscopic graphite, $\mu_C^L = g_G^{\text{bulk}}$. Define

$$S_G \equiv (a_C)/(a_C^{\text{sat},G}) = \exp[(\mu_C^L - g_G^{\text{bulk}})/(RT)]. \qquad \text{(S10)}$$

This makes $S_G=1$ the bulk graphite reference. Values above one make completed-repeat incorporation into bulk graphite downhill; values below one favor bulk graphite dissolution.

The corresponding pure macroscopic diamond completed-repeat threshold on the graphite-normalized scale is

$$S_{G,D}^{\text{eq}}(T,P) \equiv \exp[(g_D^{\text{bulk}} - g_G^{\text{bulk}})/(RT)]. \qquad \text{(S11)}$$

A diamond-normalized activity would instead be $S_D \equiv \exp[(\mu_C^L - g_D^{\text{bulk}})/(RT)] = S_G/S_{G,D}^{\text{eq}}$. It equals one at the diamond repeat threshold. To keep all comparisons on the graphite-normalized scale, this Supporting Information uses $S_{G,D}^{\text{eq}}$ for the threshold and does not use $S_D$ as its name.

The completed-repeat event free energies are

$$\Delta Y_G^{\text{rep}} = -RT\ln S_G, \qquad \Delta Y_D^{\text{rep}} = RT\ln((S_{G,D}^{\text{eq}})/(S_G)). \qquad \text{(S12)}$$

At the low pressures and temperatures considered here, the bulk ordering gives $S_{G,D}^{\text{eq}} > 1$. For $S_G < 1$, both repeats are uphill; at $S_G = 1$, macroscopic graphite is at coexistence; if pressure or temperature made diamond the stable bulk phase, $S_{G,D}^{\text{eq}}$ would be less than one. For $1 < S_G < S_{G,D}^{\text{eq}}$, graphite incorporation is downhill but diamond incorporation remains uphill. For $S_G > S_{G,D}^{\text{eq}}$, both are downhill. Equation S12 does not predict nucleation or retained product.

A metal can shift the composition or gas feed corresponding to a chosen $S_G$ by changing activity coefficients. It cannot reverse the ordering in Equation S12 for pure macroscopic graphite and diamond. Metastable phase diagrams that show a diamond field in a metal–carbon system may be computed by suppressing graphite, as is done for Fe–C in ref 25. Such a field shows where diamond would be stable if graphite were absent. It does not show that diamond is selected when graphite is available, which is the question addressed here.

## S2.2. Why the diamond threshold is reported as "about 2.0 at 1300 K"

**Purpose.** This subsection makes the thermochemical benchmark reproducible and states its uncertainty transparently.

Define $\Delta g_{DG}(T,P) \equiv g_D^{bulk} - g_G^{bulk}$. White and co-workers measured $\Delta g_{DG}$ at 298.15 K and ambient pressure as +3.170 ± 0.150 kJ mol$^{-1}$; their reported diamond → graphite value has the opposite sign.[23] Fried and Howard provide explicit $G(P,T)$ functions for graphite and diamond over a broad modeled interval. Their diamond heat-capacity comparison is experimentally constrained to approximately 1200 K, so evaluation at 1300 K is a modest model extrapolation, not a direct measurement.[24]

The calculation retains the newer 298.15 K calorimetric anchor and uses the Fried-Howard model only for the temperature increment:

$$\Delta g_{DG}^{anch}(T)=\Delta g_{DG}^{White}(298.15\,\mathrm{K})+[\Delta g_{DG}^{FH}(T)-\Delta g_{DG}^{FH}(298.15\,\mathrm{K})]. \qquad \text{(S13)}$$

Direct evaluation of the Fried-Howard model gives approximately 2.72 kJ mol$^{-1}$ at 298.15 K and, at 1300 K, approximately $\Delta g_{DG}$=6.78 kJ mol$^{-1}$ and $S_{G,D}^{eq}$=1.87. Reanchoring its temperature increment to the White value gives approximately $\Delta g_{DG}$=7.23 kJ mol$^{-1}$ and $S_{G,D}^{eq}$=1.95. Both figures inherit a reference-state parameter that requires comment.

Of the Fried-Howard increment of 4.06 kJ mol$^{-1}$ between 298.15 and 1300 K, 3.05 kJ mol$^{-1}$ arises from the term $-(T - T_0)$ [$\Delta S_0$(diamond) − $\Delta S_0$(graphite)] and 1.01 kJ mol$^{-1}$ from the Einstein heat-capacity integrals. The reference entropy that dominates the increment is $\Delta S_0$(diamond) = 2.70 J mol$^{-1}$ K$^{-1}$, whereas the accepted standard entropy of diamond is 2.377 J mol$^{-1}$ K$^{-1}$. Substituting the accepted value raises the increment to 4.38 kJ mol$^{-1}$, giving $\Delta g_{DG}$ = 7.55 kJ mol$^{-1}$ and $S_{G,D}^{eq}$ = 2.01 at 1300 K. This shift, 0.32 kJ mol$^{-1}$ or 0.06 in the threshold, exceeds the propagated experimental uncertainty of the 298 K anchor.

An independent route agrees with the corrected value. The assessed Gibbs-energy functions for graphite and diamond are printed as Equations 11 and 12 of Gustafson's evaluation of the Fe–C system,[25] and are taken from his evaluation of carbon.[26] Each of those two equations also carries a term $\int_0^P V_m\,dP$. I omit it. At 1298 K and 1 atm it adds 0.55 J mol$^{-1}$ to graphite and 0.35 J mol$^{-1}$ to diamond, so 0.20 J mol$^{-1}$ to the difference, and $S_{G,Deq}$ changes by $4 \times 10^{-5}$. Gustafson states that these functions describe the available data between 298 and 5000 K and between 0 and 15 GPa, and that extrapolation is reasonable to 6000 K and 40 GPa. The evaluation at 1298 K therefore lies inside the range he assessed. Integrating their full heat-capacity expressions, which share neither the ambient anchor nor the temperature model used above, gives $\Delta g_{DG}$ = 7.36 kJ mol$^{-1}$ at 1298 K and $S_{G,D}^{eq}$ = 1.98. The same integration reproduces $\Delta_f H°$ = +1,895.8 J mol$^{-1}$ and $\Delta_f S°$ = −3.383 J mol$^{-1}$ K$^{-1}$ at 298.15 K, within 0.6% of tabulated values, and over 900–1500 K the result is reproduced to within 0.002 by $S_{G,D}^{eq}(T) = \exp(151.4/T + 0.5651)$. Naraghi and co-workers re-assessed Fe–C in 2014 and kept both of these functions unchanged above 298.15 K.[27] They print the leading constants as −17,368.441 and −16,359.441 J mol$^{-1}$. Each differs from Gustafson's rounded value by 0.559 J mol$^{-1}$, and that shift cancels in the difference. What they changed lies below 298.15 K, where they added Einstein terms to carry the functions down to 0 K. I did not survey assessments published after 2014, so this shows only that the description was current at that date. At 298.15 K the same functions give ΔgDG = 2.904 kJ mol−1, which is 0.27 kJ mol−1 below the White anchor and outside its stated ± 0.150. The two routes therefore agree at 1300 K while disagreeing about the ambient gap by more than the anchor's own uncertainty. The agreement at 1300 K follows from compensating temperature increments and is not independent confirmation of the 298.15 K value. The two descriptions that do not depend on the Fried-Howard reference entropy therefore give 1.98 and 2.01; the working value adopted here is about 2.0, and the descriptions examined span approximately 1.95 to 2.01.

Two notes for a reader reproducing these figures. Equation (7) of the Fried-Howard paper contains a typographical error: the printed Einstein entropy term $-\ln(1 - e^{x})$ is undefined for $x = \Theta/T > 0$ and should read $-\ln(1 - e^{-x})$; the corrected form reproduces the paper's own $S_0$ when its $C_{p,0}(T)/T$ is integrated numerically. The anchor value is quoted from Table 1 of White *et al.*; their abstract and conclusion relabel the same quantities as formation values with inconsistent signs, so Table 1, or Table S19 of their Supporting Information, should be used.

The uncertainty in the White anchor (its stated ± 0.150 treated as a one-standard-deviation uncertainty) propagates to the logarithm of the threshold as

$$\sigma[\ln S_{G,D}^{eq}]=(\sigma_{\Delta g})/(RT). \qquad \text{(S14)}$$

At 1300 K, propagation of the 298 K experimental anchor alone gives $S_{G,D}^{eq} \approx$ 1.925–1.979 when the unrounded White anchor, 3.1744434 kJ mol$^{-1}$, is used. This is an anchor-only sensitivity range, not a 95% confidence interval for the 1300 K threshold, because the pressure correction and graphite reference-state uncertainties have not been quantified. The appropriate wording is therefore "about 2.0 at 1300 K with the stated thermochemical data and model."

**Table S3. Quantities used in the 1300 K benchmark.**

| Description | $\Delta g_{DG}$ (kJ mol$^{-1}$) | $S_{G,D}^{eq}$ | Note |
|---|---|---|---|
| White 298.15 K anchor, $\Delta g_{DG}$ | 3.170 ± 0.150 | — | Experimental ambient-pressure value at |

| Description | $\Delta g_{DG}$ (kJ mol$^{-1}$) | $S_{G,D}^{eq}$ | Note |
|---|---|---|---|
| | | | 298.15 K; not a 1300 K measurement |
| Direct Fried-Howard model at 1300 K | ≈ 6.78 | ≈ 1.87 | Model evaluation; uses $\Delta S_0$(diamond) = 2.70 J mol$^{-1}$ K$^{-1}$ |
| White-anchored Fried-Howard increment | ≈ 7.23 | ≈ 1.95 | Working data-and-model choice; shares that reference entropy |
| Assessed heat-capacity integration at 1298 K | ≈ 7.36 | ≈ 1.98 | Functions as printed in refs 25 and 26 and retained in ref 27; independent temperature model, sharing neither the anchor nor the reference entropy. Gives 2.904 kJ mol−1 at 298.15 K, below the White anchor |
| Fried-Howard increment with the accepted $\Delta S_0$(diamond) | ≈ 7.55 | ≈ 2.01 | Uses 2.377 in place of 2.70 J mol$^{-1}$ K$^{-1}$ |
| White-anchor propagation only | — | ≈ 1.925–1.979 | Sensitivity to the ± 0.150 anchor alone, using the unrounded 3.1744434 kJ mol$^{-1}$ value; narrower than the spread across descriptions and not a full confidence interval |
| Reported estimate | — | About 2.0 | Descriptions examined span approximately 1.95–2.01 |

### S2.3. A finite graphite object has its own marginal boundary

**Purpose.** Bulk graphite and a small, edge-rich graphite object are not thermodynamically identical. The finite-object boundary must include reservoir work associated with non-carbon species and must follow the actual state-evolution path.

To avoid confusion with the saturation $S_G$, denote the full graphite state by $\chi_G$. It includes layer number or thickness, edge orientation and reconstruction, termination, support, strain, defects, passivation, interfacial state, and every non-carbon inventory allowed to exchange with reservoirs. Let $\lambda$ denote a declared physical growth/dissolution path $\chi_G^{\lambda}(N_C)$ at stated $M$, temperature, pressure, support, and external controls.

Let $G_G^{decl}$ denote the constrained reversible free energy already appropriate to the declared mechanical and electrical ensemble, including any required Legendre transforms. For the purpose of adding or removing carbon, define the carbon-partially reservoir-transformed free energy

$$Y_G^{(C)}(N_C,\chi_G,M)=G_G^{decl}-\sum_{i\neq C} N_i\mu_i^{res}. \qquad (S15)$$

Equation S15 leaves carbon untransformed so that the derivative with respect to $N_C$ is the reversible carbon-transfer cost. If a mechanical or electrical work coordinate is not already included in $G_G^{decl}$, its ensemble-specific work or Legendre term must be added explicitly with the correct sign; a generic $-X\xi$ term is not assumed.

The marginal finite-graphite carbon potential along path $\lambda$ is

$$\mu_{C,G}^{finite,\lambda}(N_C,\chi_G,M)=(dY_G^{(C)}[N_C,\chi_G^{\lambda}(N_C),M])/(dN_C). \qquad (S16)$$

This is a **total derivative along the declared path**. If non-carbon inventories exchange with reservoirs, or if layer number, edge state, support contact, passivation, or reconstruction changes with $N_C$, those changes are included through Equation S15 and $\chi_G^{\lambda}(N_C)$. A partial derivative at fixed $\chi_G$ is appropriate only for a genuinely fixed state family. Where the carbon amount changes discretely, or the declared path crosses a reconstruction, layer-number, or termination transition, the corresponding one-growth-unit finite difference of Equation S15 replaces the derivative.

When non-carbon inventories and generalized-work coordinates are fixed, Equation S15 can be written $Y_G^{(C)}=N_C g_G^{bulk}+G_{ex}$, and Equation S16 reduces to $g_G^{bulk}+dG_{ex}/dN_C$ along the path.

The graphite-normalized reversible boundary of the declared object is

$$S_G^{finite,eq,\lambda}(N_C,\chi_G,M)=\exp[(\mu_{C,G}^{finite,\lambda}-g_G^{bulk})/(RT)]. \qquad (S17)$$

For a macroscopic diamond seed, let $S_G^{\mathrm{op}}$ denote the operating graphite-normalized activity. A **pairwise finite-competitor interval** exists only when

$$S_{G,D}^{\mathrm{eq}} < S_G^{\mathrm{op}} < S_G^{\mathrm{finite,eq},\lambda}(N_C,\chi_G,M). \qquad \text{(S18)}$$

Equation S18 compares the seed with one declared graphite state and path. It does not suppress all graphite states, bulk graphite nuclei, amorphous carbon, carbides, or wall deposits. Those competitors require separate state and rate bounds. The interval cannot be inferred from finite-time graphite nondetection; the finite object must be present, controlled, and tested reversibly. The superscript eq presumes a reversibly traversed declared path; growth- and dissolution-branch hysteresis must be excluded (Protocol 4) before the boundary is treated as an equilibrium property.

## S2.4. Why the two-dimensional edge correction contains a factor of one-half

**Purpose.** This derivation distinguishes the mean edge excess from the marginal contribution that controls a reversible feature-growth boundary and corrects the basal-area benchmark.

For an isotropic self-similar two-dimensional feature, let $E_{\mathrm{edge}}=\beta\mathcal{P}$ and $A=N_C\bar{A}_C$, where $N_C$ is the amount of carbon, in moles, and $\bar{A}_C$ is basal area per mole of C. Here, $\mathcal{P}$ denotes perimeter length; in equations and prose, P denotes pressure. Self-similar growth gives $\mathcal{P}\propto A^{1/2}$. Therefore

$$\delta\mu_{\mathrm{edge}}=(\partial E_{\mathrm{edge}})/(\partial N_C)=(\beta\bar{A}_C\mathcal{P})/(2A)=(\beta\bar{A}_C)/(2\ell), \qquad \ell\equiv(A)/(\mathcal{P}). \qquad \text{(S19)}$$

The mean edge excess $E_{\mathrm{edge}}/N_C=\beta\bar{A}_C/\ell$ is exactly twice the marginal value for this restricted self-similar case. Equation S19 uses line free energy $\beta$, not step stiffness $\tilde{\beta}$. If shape, reconstruction, passivation, support contact, or $\beta$ changes with size, the total derivative in Equation S16 must be used.

Neutron diffraction gives an in-plane graphite C–C distance of 0.1422 nm.[28] Honeycomb geometry then gives the basal area per carbon atom:

$$A_C=(3\sqrt{(3)})/(4)d_{\mathrm{C-C}}^2=0.0263\ \mathrm{nm}^2\ \text{per C atom.}$$

Using the illustrative $\beta$ = 1 eV nm$^{-1}$ and $\ell$ = 5 nm, Equation S19 gives 0.00263 eV per C atom, or 0.253 kJ mol$^{-1}$. Matching the 7.55 kJ mol$^{-1}$ working diamond–graphite gap with the edge term alone would require $\ell \approx$ 0.168 nm. Here $\ell = A/\mathcal{P}$ is the area-to-perimeter ratio, not a feature width. It corresponds to a circle diameter or square side of about 0.67 nm, or an equilateral-triangle side of about 1.16 nm. These dimensions are atomistic, where the continuum separation between perimeter and interior is not quantitative. The simple edge-only model consequently fails before it can close the bulk gap. No universal crossover length is claimed, and $\ell$ = 5 nm is used here only as an illustrative feature scale.

Figure 3 of the main text uses three explicitly labeled sensitivity values, $\beta$=0.5, 1, and 2 eV nm$^{-1}$. They are illustrative inputs, not a measured uncertainty interval for the relevant metal-supported graphite edge.

For a smooth isotropic three-dimensional interface, the corresponding curvature shift is

$$\delta\mu_{\mathrm{curv}}=\gamma\bar{V}_u\kappa. \qquad \text{(S20)}$$

Here $\kappa$ is the sum of principal curvatures with an outward normal and positive curvature for a convex particle; for a sphere, $\kappa = 2/r$. Equation S20 is not valid unchanged for a strongly faceted, anisotropic, stressed, charged, or compositionally segregated interface.

## S2.5. Search criterion for a finite-graphite pairwise window

**Purpose.** This subsection makes explicit an asymmetry that is implicit in Equations S15–S19, and converts it into a concrete target for experiment and modeling.

The two sides of the pairwise comparison are not symmetric. A seeded diamond front advancing by completed translational repeats pays the bulk free energy per growth unit. A finite graphite object can instead have a state-dependent additional marginal free-energy cost for the next carbon increment along its declared path $\chi_G(N_C)$. The pairwise interval of Equation S18 exists only if this marginal cost exceeds the bulk graphite value by at least the bulk diamond-graphite gap—approximately 7.55 kJ mol$^{-1}$ at 1300 K under the stated data and model, or about 0.078 eV per carbon atom—over the relevant carbon-content range.

Section S2.4 shows that the declared continuum edge term does not supply this shift at the illustrative scale: for $\ell = 5$ nm and $\beta = 1$ eV nm$^{-1}$, the marginal edge contribution is approximately 0.25 kJ mol$^{-1}$. A state that opens the interval could instead involve reconstruction, termination, metal passivation, support chemistry, layer discreteness, strain, curvature, defects, or other state-specific effects. Each contribution can be calculated or measured. The edge estimate establishes none of them.

This gives a specific search criterion. Modeling should screen declared terminated, metal-contacted, supported, or strained graphite states under the reservoir conditions of the chosen flux and rank their marginal carbon potentials above $g_G^{bulk}$. Experiment should prepare accessible high-ranked states and measure their reversible carbon-transfer boundaries (Protocol 4). Both outcomes are informative, although only a confirmed boundary above the diamond-front threshold is decisive for that pair. Failure among the states examined would support, but not establish, kinetic suppression of graphite.

This screening will be demanding. GPUMD demonstrates large-scale atomistic simulation, and the cited TorchNEP preprint concerns efficient potential training.[29,30] Neither validates a potential for the relevant Ga-Fe-Ni-Si-C melt, interfaces, carbide states, or finite graphite states. Training and validation must cover the actual compositions and structures. Because the screen discriminates at approximately 0.078 eV per carbon atom, uncertainty for the compared free-energy differences must be demonstrated below that scale.

The declared-state requirement does not create the multiplicity of graphite boundaries; it exposes a physical multiplicity. There is no single finite-graphite boundary. Every graphite state $\chi_G$ has its own reversible boundary. A claim of “graphite suppression” is unreproducible unless it identifies the object.

For the pairwise thermodynamic comparison, the most restrictive accessible competitor is the state with the lowest reversible boundary in the chosen system. Bulk graphite provides the reference $S_G = 1$, but the boundary of a finite supported state can move in either direction: support contact, metal wetting, edge chemistry, and strain can stabilize a finite carbon state as readily as destabilize it, so its position must be calculated or measured rather than assumed. The question is whether the lowest boundary found across the competitor states the system can actually reach—graphitic, carbidic, or amorphous, with no ordering assumed in advance—lies above the diamond threshold. This comparison does not predict which state will nucleate or persist; rates and retained inventories remain separate.

Witness selection follows from this: the most relevant witnesses are replicas of the graphitic material the chosen system itself produces—matched in layer number, edge state, support, and metal contact to post-run characterization—tested in order from the most dangerous (near-bulk, lowest expected boundary) toward more highly modified states. Each measured state-boundary pair, even for a deliberately artificial witness, remains a transferable calibration point: it tests whether atom-scale effects can supply additional free-energy costs of the required magnitude at all, and it validates the models used to predict the boundaries of states not yet measured. The system-level claim that no carbon competitor accumulates rests on carbon-inventory closure (Section S3.3 and Protocol 5). No single witness can establish it. What the witness measurements add is the reason it holds.

One caveat bounds this picture. The carbonaceous states anchored at $S_G = 1$ are referenced through their marginal carbon potential. A carbide’s formation also consumes metal atoms, whose chemical potentials are set by the melt itself, so carbide stability is not referenced to graphite saturation at all: for a strongly carbide-forming component, carbide can be a downhill carbon sink even below $S_G = 1$. In carbide-forming fluxes—the related Ga–Si seeded system being the experimental example cited here—the operative worst-case competitor may therefore be a carbide rather than any graphite state, and its boundary must be located in the appropriate two-potential space (compare the BN coordinates of Section S5).

## S2.6. Carbide competitors are referenced to the metal’s own chemical potential

**Purpose.** This subsection states why carbide competitors cannot be bounded on the graphite-normalized scale alone, supporting the caveat of Section S2.5. No new numbered relations are introduced.

For a carbide formed from dissolved carbon and a metal component of the flux, incorporation of carbon is thermodynamically downhill when the carbon chemical potential exceeds the carbide’s molar free energy per carbon after crediting the metal atoms at their own chemical potentials in the melt. The melt itself fixes those metal chemical potentials—near the pure-liquid value when the component’s activity is near unity—so a carbide’s threshold is set jointly by the carbide’s stability and the metal’s chemical potential, not by graphite saturation. On the graphite-normalized scale, the corresponding threshold can therefore lie below $S_G = 1$ for a strongly carbide-forming component: carbon uptake by the carbide is then downhill even in a melt undersaturated with respect to graphite. Gustafson describes cementite with a constant heat capacity and no magnetic term. That omission is not small. In the later description the magnetic entropy of cementite is $R\ln(\langle\beta\rangle+1) \approx 10.3$ J mol$^{-1}$ K$^{-1}$, and it does not fall off with temperature.[27] What does fall off is the short-range-order term, which is −0.35 J mol$^{-1}$ at 1300 K. Gustafson’s fitted linear term absorbs the rest over the range he optimized. The test is whether the two descriptions agree on what was measured, and they do: the metastable eutectic and eutectoid differ by one degree or less. The later work also adds

Hägg carbide (χ-$Fe_5C_2$) and eta carbide (η-$Fe_2C$), each more stable than cementite only below 720 K and 470 K. Cementite is therefore the carbide that matters at the temperatures used here. Adding accessible carbide states can only put further thresholds below the diamond threshold. $Fe_7C_3$ is in none of these three assessments, and I make no claim at the pressures where it matters. The assessed Fe–C system illustrates that ordering directly. With graphite suppressed, the calculated metastable phase diagram plotted against carbon activity places the cementite field across most of the interval between graphite saturation and the diamond field, which is entered only near the upper end of the plotted activity range.[25] Fe is one component of the demonstrated flux, but the binary parameters do not transfer to the quaternary. What transfers is the ordering of the two thresholds on a graphite-referenced activity axis, not a number.

Three consequences follow. First, in carbide-forming fluxes the lower envelope of Section S2.5 must include accessible carbide states and is not anchored at $S_G = 1$. Second, a finite carbide patch has its own finite-state costs, treated as in Equations S15–S17 with the appropriate reservoir set. Third, the operating point must lie below the reversible boundaries of accessible carbide states or carbide appearance must be bounded kinetically. In the related Ga–Si seeded system described in a non-peer-reviewed manuscript deposited in ChemRxiv, SiC particles and epitaxial 3C-SiC on diamond were reported, and Si participation in the front chemistry was proposed.[31] These observations establish that SiC formed alongside diamond in that Ga–Si system. They motivate measurements of carbide state and rate bounds for the Ga–Fe–Ni–Si reservoir but do not establish those bounds.

The same ChemRxiv manuscript illustrates strong Si sensitivity and carbide coexistence. It does not establish an equilibrium margin.[31] At 0.57 at% Si, dense aligned diamond pyramids formed and SiC particles were observed between dense regions. At 0.62 at% Si, both SiC and pyramidal diamond were reported. SiC was therefore present at both compositions; the two qualitative runs do not locate a 0.05 at% threshold or establish phase fractions. In two imaged examples, 3C-SiC at pyramid apices was lattice-aligned with the underlying diamond. The observations show that Si concentration affected the deposits and that carbide formation was active under both conditions. A reversible boundary, activity model, replicate series, and phase-resolved inventory are still required. The same source notes that the carbide may itself be consumed during diamond growth and reports pyramid apices free of it, so persistence of the carbide sink is not established either.

## S2.7. Candidate condensed carbon sources

This subsection identifies condensed carbon sources worth testing and distinguishes their nominal thermodynamic state from the carbon potential delivered to the seed.

The state of a condensed carbon source does not by itself determine the carbon chemical potential at the diamond seed. Carbon must leave the source, enter the liquid, move through it, and reach the seed without being consumed by graphite, amorphous carbon, carbides, deposits elsewhere in the reactor, or volatile products. The source and seed need not be at the same temperature. A controlled temperature gradient could separate source dissolution or decomposition from incorporation at the seed, but the resulting chemical-potential and concentration profiles must be measured or modeled for the chosen liquid.

A $^{13}$C-enriched diamond seed permits candidate sources to remain at natural isotopic abundance. A $^{12}$C-rich diamond layer extending outward from the registered surface of the original $^{13}$C-rich seed would show incorporation of external carbon during net diamond advance. The isotope boundary must be interpreted together with registered interface displacement, crystallographic continuity, seed recession, lattice exchange, surface roughness, diffusion, depth resolution, and isotope fractionation. All other natural-abundance carbon reservoirs—including crucibles, fixtures, binders, pre-existing deposits, and adventitious carbon—must be removed or independently bounded. A $^{13}$C-rich seed distinguishes seed carbon from external carbon, but not one natural-abundance external reservoir from another.

I would begin with natural-abundance bulk diamond in a controlled temperature gradient. It is a crystalline, compositionally well-defined source and avoids introducing a graphite-source interface. Its dissolution may be slow, and a bulk diamond source at the same temperature and pressure as a macroscopic diamond seed cannot by itself supply a carbon potential above the diamond equilibrium value. A temperature difference, source curvature, stress, defects, or another independently established contribution is therefore required to drive net transport from source to seed. I would next test natural-abundance nanodiamond. Its curvature, defects, and large surface area may raise its free energy and accelerate carbon release, but aggregation, particle transport, surface termination, oxygen-containing groups, and direct incorporation of particles require separate controls.[65]

Glassy carbon provides a useful warning against assigning one free energy to a broad material class. Das and Hucke measured the carbon activities of several specified vitreous carbons relative to macroscopic graphite. Their most favorable specimen, ZC-23, was prepared from Union Carbide 321-13 material and heat-treated at 2000 °C under nitrogen. Its reported carbon activity was 4.34 at 600 °C and 1.22 at 1000 °C. Values calculated from the authors' reported enthalpy and entropy are approximately 2.05 at 800 °C and 1.55 at 900 °C.[66]

For comparison, the graphite-normalized completed-repeat threshold for pure, unstressed macroscopic diamond near 1 atm is approximately 2.09 at 600 °C, 2.03 at 800 °C, 2.00 at 900 °C, and 1.98 at 1000 °C under the thermochemical description adopted in Section S2.2. The 600 °C threshold is an extrapolation of the compact fit below its stated 900–1500 K reproduction interval. ZC-23 therefore presents an interesting example of a solid carbon whose activity at lower temperature can substantially exceed graphite saturation and may reach the diamond threshold. At 800 °C, however, its calculated activity exceeds the calculated diamond threshold by only about one percent. The rounded historical thermodynamic quantities and their uncertainties do not support treating this small difference as proof that ZC-23 can drive diamond growth.

The same specimen shows how strongly the carbon activity of a metastable disordered carbon can change with increasing temperature near atmospheric pressure. Its activity decreases from 4.34 at 600 °C to 1.22 at 1000 °C, whereas the calculated diamond threshold decreases only from approximately 2.09 to 1.98. The Gibbs-energy difference contains both enthalpic and configurational-entropic contributions, and the latter becomes increasingly important as temperature rises. A carbon that is highly active at 600–800 °C can therefore fall below the diamond threshold and eventually below graphite saturation as its temperature increases. This temperature dependence also suggests a source–seed gradient experiment in which a relatively cool, high-activity carbon supplies carbon through the liquid to a seed held under separately optimized conditions. Whether that carbon potential survives source dissolution, transport, competing reactions, and delivery to the seed must be determined experimentally.

The result is not general to glassy carbon. Beckwith D-82-2 vitreous carbon, gas-baked at 2000 °C, had reported activities of 0.10 at 600 °C and 0.04 at 1000 °C. Hercules H-54-derived glassy carbon, treated at 1795 °C in vacuum, had corresponding activities of 0.05 and 0.02. Other specimens in the same study also differed substantially. “Glassy carbon” therefore does not specify a carbon chemical potential. The precursor, commercial designation, preparation procedure, maximum heat-treatment temperature, atmosphere, and thermal history must be stated. Historical commercial designations do not establish that the same material remains available or that a currently marketed grade has the same thermodynamic properties.[66]

High-purity, well-characterized glassy carbon is possible. Takahashi and Westrum measured the low-temperature thermodynamic properties of Tokai grade GC-30S after heat treatment at 3000 °C. The reported metallic impurities were in the low-parts-per-million range, although the specimen contained 0.2 ± 0.1 wt% oxygen. This study shows that a particular glassy-carbon grade can be defined by composition and thermal history and can have reproducibly measured thermal functions. It does not assign the same thermodynamic state to all glassy carbons. For a large piece, surface functional groups make a negligible contribution to the bulk composition and bulk free energy. They can nevertheless affect the initial liquid–solid reaction because carbon transfer begins at the surface.[67]

I would also test C60 because it is a molecular solid containing only carbon and has a measured formation enthalpy substantially above graphite. Its high nominal energy does not establish the activity reaching the seed through the liquid: C60 may polymerize, fragment, dissolve as another carbon species, or form disordered or graphitic carbon before carbon reaches the seed. The products recovered at the source, throughout the liquid, and at the seed must therefore be identified.[68]

Selected carbides would follow only when their reactions with the chosen liquid provide a carbon potential that can be calculated, adjusted, and experimentally calibrated. Stable carbides can instead consume carbon and lower its activity, while highly metastable carbides may decompose before establishing a useful source condition. Any released metal can also change the liquid composition, activity coefficients, interfacial state, and competing-phase kinetics. Section S2.6 gives the required two-potential treatment. Ferrocene likewise belongs later in the sequence because it supplies Fe together with carbon-containing fragments; its carbon-delivery function must be separated from the effect of Fe.[69]

Polymers and molecular hydrocarbons—including polyethylene, polystyrene, carbon-rich aromatic solids, and halogenated hydrocarbons—could eventually be examined as carbon-bearing precursors. During heating and contact with a forming or fully liquid metal flux, they may undergo several competing reactions and may also produce volatile species or deposits elsewhere in the reactor. The metal composition, heating sequence, source placement, gas environment, and temperature field could each change these pathways. Halogenated precursors introduce additional possibilities through reactions with hydrogen, the liquid, the alloying elements, the reactor materials, and the diamond surface. The nominal composition and free energy of the starting material therefore do not specify the carbon potential delivered to the seed. Such precursors may ultimately prove useful, but the many coupled gas-phase, liquid-phase, interfacial, and competing-carbon pathways would make the origin and meaning of any observed diamond growth difficult to establish. I would pursue them only after better-defined condensed sources have been tested and the activity, transport, isotope-registration, and carbon-inventory measurements are in place.

As a provisional experimental order, I would first test bulk diamond under a controlled temperature gradient, followed by nanodiamond with its surface chemistry and particle transport controlled. I would then examine a reproduced ZC-23-type resin-derived carbon in the lower-temperature interval in which its reported activity is favorable, followed by other high-purity glassy or amorphous carbons whose activities have been calibrated against graphite. C60 and selected carbides would follow.

Ferrocene requires experiments that separate carbon delivery from the effect of Fe. Polymers and other molecular hydrocarbons would be placed later because their pathways during heating and contact with the forming or fully liquid flux are not yet defined. This provisional ordering is based on educated guesses about which experiments are most likely to give interpretable results; it is not a predicted ranking of diamond yield. As experiments and modeling proceed, the ordering may of course change. Every series requires graphite calibration, registered outward growth on a $^{13}C$-enriched seed, and phase-resolved carbon accounting.

## S3. From event thermodynamics to what an experiment observes

### S3.1. A downhill event can still be slow

**Purpose.** The sign of $\Delta Y$ determines thermodynamic direction. The barrier additionally requires a declared reaction coordinate or dividing surface.

For pathway $\alpha$, let $q$ be a reaction coordinate connecting the declared initial and final basins, and let $q_\alpha^\ddagger$ denote the dividing surface or maximum of the constrained reversible free-energy profile. Define

$$\Delta Y_\alpha^\ddagger = Y(q_\alpha^\ddagger) - Y(s).$$

A schematic activated-event or nucleation rate density is

$$\mathcal{J}_\alpha = \mathcal{J}_{0,\alpha} \exp(-(\Delta Y_\alpha^\ddagger)/(RT)). \qquad \text{(S21)}$$

The molar and per-event conventions are equivalent when $\Delta Y_\alpha^\ddagger = N_A \Delta G_{\alpha,\text{event}}^\ddagger$, so $\Delta Y_\alpha^\ddagger/RT = \Delta G_{\alpha,\text{event}}^\ddagger/k_B T$. The event basis must be stated. $J_{0,\alpha}$ is an effective prefactor on that same declared exposure basis; it is not an attempt frequency by itself. The prefactor contains attempt frequency, concentrations, transport, transmission coefficient, and site-population factors. A lower final-state free energy does not imply a lower barrier.

An Arrhenius plot made while temperature also changes carbon activity, dissolved-carbon speciation, transport, available sites, or the front state gives an apparent activation energy. Its slope cannot by itself be assigned to $\Delta Y_\alpha^\ddagger$ or to a unique activation enthalpy. Cross-temperature comparisons should either hold fixed the declared event affinity and verify the front state, or fit the joint dependence on temperature and activity, while accounting for the prefactor and transport.

For statistically independent events, let $d\mathcal{E}(t)$ be the differential active exposure—volume-time, area-time, site-time, or another declared basis—and let $\varepsilon_\alpha(t)$ be the event-detection/recovery efficiency. Define the integrated detected-event hazard $H_\alpha$ and the probability $\Pr_\alpha$ of at least one detected event by

$$H_\alpha \equiv \int \varepsilon_\alpha(t) \mathcal{J}_\alpha(t)\, d\mathcal{E}(t), \qquad \Pr_\alpha = 1 - \exp(-H_\alpha). \qquad \text{(S22)}$$

For constant $\mathcal{J}_\alpha$, constant efficiency $\varepsilon$, total active exposure $\mathcal{E}$, and zero observed events, the conventional one-sided 95% Poisson upper bound on the rate is

$$\mathcal{J}_{\alpha,95} = -\ln(0.05)/(\varepsilon\mathcal{E}). \qquad \text{(S23)}$$

If efficiency or active exposure is uncertain, that uncertainty belongs in the likelihood model. “No graphite detected” is therefore a rate bound that depends on the protocol. It carries no equilibrium information.

### S3.2. Zero interface velocity is not automatically zero affinity

**Purpose.** A stationary interface can result from reversible balance, balanced independent processes, or kinetic arrest. These cases must be distinguished.

Let $j_{D,k}^{\text{att}}$ and $j_{D,k}^{\text{det}}$ be the forward attachment and reverse detachment fluxes of the same diamond-kink reaction on a common carbon, area, and time basis. Let independent losses describe etching, conversion to graphite, or other removal pathways. The retained diamond flux and normal interface velocity are

$$j_D^{\text{ret}} = j_{D,k}^{\text{att}} - j_{D,k}^{\text{det}} - j_{\text{etch}} - j_{D\to G} - j_{\text{other}}, \qquad v_n = \bar{V}_D j_D^{\text{ret}}. \qquad \text{(S24)}$$

At the true completed-repeat zero affinity, the matched pair satisfies $j_{D,k}^{\text{att}} = j_{D,k}^{\text{det}}$. An observed $v_n=0$ coincides with that condition only when independent gain and loss channels are absent or quantitatively subtracted.

A zero-velocity plateau may instead reflect passivation, pinning, a two-dimensional nucleation delay, or insufficient displacement sensitivity. Likewise, pure macroscopic facets have the same completed-repeat threshold but can have different step densities, mobilities, barriers, stresses, curvatures, terminations, and finite-time onsets. Vacuum, electrochemical, solution, diamond-dissolution, and diamond step-flow studies show that relevant step, kink, boundary, and morphology observables can be measured.[32–37]

Solution growth of elemental semiconductors supplies a measured instance of step-source exhaustion. A schematic comparison places representative liquid-phase epitaxy near 0.1–1% supersaturation.[38] Under cited near-equilibrium, faceted, dislocation-free conditions, growth proceeded by existing steps and stopped after the surface closed on stable low-index facets and the step source disappeared; the reservoir remained supersaturated.[39] This is a positive bulk reservoir driving force without an available step event. On an amorphous substrate, failure to nucleate a crystal interface does not define the affinity of a nonexistent growth event. For diamond, a stationary faceted seed therefore does not locate the repeat boundary unless step availability is independently verified.

Solution growth of SiC supplies a second instance, in which a transient rather than an exhausted step source produced the zero crossing. An in situ confocal laser scanning microscope focused through a semi-insulating SiC seed tracked individual steps at four frames per second, with a minimum resolvable step height of about 6 nm.[40] In a Si-Cr solvent held at 1750–1800 °C, the heating laser was stopped. The steps reversed near 1800 °C, receded for about 300 s, and advanced again near 1600 °C.[41] In a pure Si solvent under the same procedure the steps slowed and stopped without reversing. The authors attribute the reversal to a lower activation energy for dissolution after Cr addition, with the elemental crystallization rate falling faster than the elemental dissolution rate while the imposed thermal gradient decayed. The carbon potential was not calibrated, the temperature was not held, and the interface was not approached from a carbon-poor and a carbon-rich side at fixed conditions. The interface velocity therefore crossed zero twice without locating a reversible boundary. This apparatus could be used for Protocol 2, but reversal of the front alone does not locate a boundary.

## S3.3. Count every carbon sink on one basis

**Purpose.** Selectivity, retained product, and feed yield are different quantities. They can be compared only when every carbon inventory uses the same control volume and time interval.

A complete carbon balance in rate form is

$$\dot{N}_C^{\mathrm{in}} - \dot{N}_C^{\mathrm{out}} = (d)/(dt)(N_C^{L} + N_C^{D} + N_C^{G} + N_C^{\mathrm{carbide}} + N_C^{\mathrm{amorph}} + N_C^{\mathrm{walls}} + N_C^{\mathrm{other}}). \qquad \text{(S25)}$$

Here the dotted inlet and outlet quantities are molar carbon flow rates; all inventory terms inside the derivative are amounts on the same control-volume basis. Unassigned carbon means the balance is not closed, and nothing more.

Define $\Delta N_C^{p,\mathrm{form}} \geq 0$ as the gross carbon inventory newly formed and retained in solid phase $p$ relative to matched blanks and the declared initial inventory. Dissolution and destruction are reported separately as signed fluxes. Retained-solid diamond selectivity is

$$f_D^{\mathrm{ret}} = (\Delta N_C^{D,\mathrm{form}})/(\Delta N_C^{D,\mathrm{form}} + \Delta N_C^{G,\mathrm{form}} + \Delta N_C^{\mathrm{carbide,form}} + \Delta N_C^{\mathrm{amorph,form}} + \Delta N_C^{\mathrm{other\ solid,form}}). \qquad \text{(S26)}$$

All denominator terms are nonnegative gross retained formation amounts on the same basis. This prevents graphite dissolution from artificially increasing selectivity above one. If no solid phase forms, the denominator is zero and $f_D^{\mathrm{ret}}$ is undefined rather than zero.

Feed-attributed retained-diamond yield is

$$\eta_D^{\mathrm{feed}} = (\Delta N_C^{D,\mathrm{form},feed})/(N_C^{\mathrm{feed}}). \qquad \text{(S27)}$$

The numerator in Equation S27 is the newly formed retained diamond carbon attributed to the feed by isotope or another validated source-allocation model. Without source attribution, the ratio is not a feed yield unless the feed is established as the sole carbon source. The quantity is undefined when no feed carbon enters the declared interval. A run can have high retained-solid diamond selectivity but low feed-attributed yield if most feed carbon remains dissolved, exits, or forms other products.

## S3.4. What isotope profiles establish

**Purpose.** Isotope labeling identifies carbon provenance, but source attribution, exchange, and net outward growth are distinct claims.

The local $^{13}$C f$\boldsymbol{r}$action at position r is

$$x_{13}(\boldsymbol{r})=(N_{13}(\boldsymbol{r}))/(N_{12}(\boldsymbol{r})+N_{13}(\boldsymbol{r})). \qquad \text{(S28)}$$

For the same control volume, isotope-resolved conservation requires

$$N_{13}^{\text{in}}-N_{13}^{\text{out}}=\Delta N_{13}^{D}+\Delta N_{13}^{G}+\Delta N_{13}^{\text{carbide}}+\Delta N_{13}^{\text{amorph}}+\Delta N_{13}^{L}+\Delta N_{13}^{\text{walls}}+\Delta N_{13}^{\text{other}}. \qquad \text{(S29)}$$

Isotope evidence should be interpreted as a ladder:

**Table S4. Minimum interpretation of isotope and interface observations. The final two entries extend the same accounting to the graphite observations.**

| Observation | What it establishes | What remains unproven |
|---|---|---|
| No detectable mass or height change | Apparent zero net change over the stated time and detection limit | Live exchange, zero local velocity, or zero affinity |
| Isotope turnover at no net mass change | Active bidirectional exchange under the stated registration, resolution, and artifact controls | Zero affinity unless the same boundary is approached from both sides and transport and independent loss terms are absent or quantitatively corrected |
| Isotope within structurally confirmed diamond | Reservoir isotope entered diamond | Net outward overgrowth rather than dissolution-reprecipitation within the original seed volume |
| Crystallographically continuous isotope-rich diamond outside a registered original interface | Net reservoir-fed seeded overgrowth with the Protocol 3 registration, resolution, blank, cleaning, and source-attribution controls | Whole-reactor selectivity and feed yield without competitor and mass balances |
| Reversible sign change of a controlled graphite feature under a small, bracketed change in the imposed condition | A pairwise growth-and-dissolution boundary for one finite object, for the declared state and the declared path | The same boundary for any other graphite state, and any statement about bulk graphite. Establishing it requires a carbon-inventory observable, tracking of the state actually present, short-time kinetics rather than an end-point comparison, and bracketing from both sides of the boundary. |
| No spontaneous graphite detected in a finite run | A protocol-dependent rate bound under the stated event-counting or hazard model—not a state boundary | No thermodynamic boundary whatever. Converting it into one requires the time-dependent exposure, a census of active sites, the detection and recovery efficiency of the method used, and a stated detection criterion. |

Before data collection, the criteria for net overgrowth should be specified. The isotope composition of the proposed new layer must be clearly distinguishable from those of the seed and blank controls after measurement uncertainties have been included. The layer must be thicker than the validated depth resolution, and its registered outward displacement must be larger than the combined uncertainty in locating the original and final interfaces. Diamond bonding, crystallographic continuity, quench and cleaning controls, phase-resolved competitor analysis, and carbon and isotope balance closure within the stated uncertainties are also required. These are proposed working criteria rather than instrument specifications. Carbon and nitrogen isotope fractionation during diamond growth has been measured and must be considered when interpreting profiles.[42–45]

## S4. Six experimental protocols that can support or falsify the proposed interpretation

Each protocol asks the same questions: What is the hypothesis? What state must be controlled? What is measured directly? What could mimic the result? What controls are required? What result counts as success? What observation would falsify the interpretation?

The protocols are written for a suitably chosen carbon-bearing metal flux *M* in contact with a diamond seed. The apparatus, activity calibration, dissolved species, and kinetic parameters are not assumed to transfer from one alloy to another.

The protocols are separate because some controls are mutually incompatible. In particular, a macroscopic graphite coupon used to establish $S_G$=1 cannot remain equilibrated with the same local reservoir during a scan to $S_G$>1.

## S4.0. Feasibility ladder and staged decision sequence

The complete program should not begin with immobilized nanographite and a fully coupled reactor. For any candidate *M*, it should proceed through the following measurable targets:

1. establish a carbon-closed equilibrium graphite reference in an inert non-carbon crucible;
2. validate at least one above-saturation activity-mapping route using multiple independent calibration data;
3. demonstrate reversible registered diamond retreat and advance on a vicinal seed at a resolvable displacement scale;
4. demonstrate isotope-attributed outward overgrowth and close the whole-reactor carbon balance;
5. begin finite-graphite tests with supported features large enough for independent state, geometry, and carbon-inventory measurements, then decrease size only after reconstruction and support reactions can be tracked;
6. measure stochastic competitors in separate runs with quantified active exposure and detection/recovery efficiency;
7. only then combine the seed and finite-competitor tests or move to a thin-film or source-to-seed scale-up geometry. This ladder is also the recommended order of attempt: each rung supplies the calibration that the next one presumes.

A co-located graphite witness is not the default because it can perturb the local carbon inventory and create heterogeneous nucleation sites. A matched separate cell is preferred. If co-location is used, the witness inventory and nucleation perturbation must be bounded relative to the melt and seed.

## S4.1. Protocol 1: macroscopic graphite calibration and activity map

**Question.** What equilibrium condition corresponds to bulk graphite saturation, and how will carbon potential above that point be quantified?

**Table S5. Protocol 1: macroscopic graphite calibration and activity map.**

| Element | Requirement |
| --- | --- |
| Hypothesis | In a carbon-closed or carbon-throughput-isolated mode, a clean macroscopic graphite reference establishes $S_G$=1 at declared *T*, *P*, chosen flux *M*, gas state, and crucible condition. An independently validated model or measurement then maps above-saturation states. |
| Controlled state | Inert non-carbon crucible; coupon grade, orientation, area, cleaning, and surface state; alloy composition; temperature field; gas state; hydrodynamics; equilibration time; and whether the carbon boundary is closed or isolated. The coupon is removed before the supersaturated scan. |
| Primary observable | Reversible coupon growth/dissolution sign, equilibrium alloy composition and coupon surface state, plus either a validated $a_C(x,T,P,M)$ model or a direct carbon-potential measurement for the chosen *M*. |
| Candidate above-saturation routes | A thermodynamic model fitted to several independent equilibria and cross-validated outside the fit; gas-liquid equilibrium against a known carbon-bearing gas reaction with verified local equilibration and conversion; or an electrochemical/EMF-type probe validated for the chosen flux, reference electrode, and temperature. |
| False positives | A nonequilibrium steady state with carbon flowing through a stationary coupon, passivation, carbide coating, incomplete equilibration, gradients, carbon loss, and treating mole fraction as activity. |

| Element | Requirement |
|---|---|
| Required controls | Approach from carbon-poor and carbon-rich sides; close or isolate carbon throughput; vary coupon area and flow after equilibration; analyze coupon and crucible surfaces; repeat at multiple temperatures/compositions; cross-validate the activity mapping against independent points. |
| Quantitative success criterion | The equilibrium reference converges from both directions within combined uncertainty and is independent of coupon area, flow, and history after equilibration. Above-saturation $S_G$ is reported only over the validated range and with model or measurement uncertainty. |
| Falsifying outcome | No unique equilibrium reference is obtained, the stationary coupon carries nonzero carbon throughput, or the activity mapping fails independent calibration data. In that case $S_G$ is not calibrated for the claimed state. |

Interpretation. A stationary coupon in an open reactor is not enough. A passing result establishes an equilibrium zero and a validated route above it.

## S4.2. Protocol 2: bracket the seeded-diamond repeat boundary

**Question.** At what validated or explicitly model-conditional carbon potential does a declared diamond front change from net retreat to net advance?

**Table S6. Protocol 2: seeded-diamond repeat boundary.**

| Element | Requirement |
|---|---|
| Hypothesis | A declared diamond front crosses from dissolution to growth as validated or explicitly model-conditional $S_G$ passes its event-specific zero-affinity condition; the pure macroscopic completed-repeat limit approaches $S_G = S_{G,D}^{eq}$. |
| Controlled state | Seed orientation, miscut, step density, termination, stress, defect density, wetted area, alloy and gas history, temperature, pressure, flow, and activity calibration. A vicinal seed supplies pre-existing steps and reduces ambiguity from two-dimensional nucleation, consistent with the BCF separation between step-flow and two-dimensional-nucleation-limited growth.[4] |
| Primary observable | Registered normal displacement and step motion versus $S_G$, cleaned mass change, isotope turnover, terrace evolution, and complete post-quench carbon inventory. |
| False positives | Balanced etching and growth, passivation, pinning, transport limitation, changing wetted area, quench deposition, stress or impurity incorporation, detector-limited apparent arrest, and exhaustion of step sources on a well-faceted, low-dislocation seed. |
| Required controls | Approach from dissolution and growth sides in small activity steps; cycle repeatedly; vary flow and miscut without changing calibrated $S_G$; include isotope exchange near zero; map stress and composition; use protected or fiducial reference regions. |
| Quantitative success criterion | A velocity sign reversal is resolved from both directions with nonoverlapping confidence intervals away from zero; the two boundary estimates agree within combined uncertainty; isotope turnover is observed at the live zero-net-growth condition; and the intrinsic boundary after stated |

| Element | Requirement |
|---|---|
| | stress, curvature, transport, and loss corrections is reproducible. |
| Falsifying outcome | No reversible sign change exists, or the corrected boundary remains dependent on hydrodynamics or observation history. For a verified pure, unstressed macroscopic front, a corrected boundary inconsistent with $S_G$=$S_{G,D}^{eq}$ falsifies the calibration or state assignment rather than demonstrating bulk diamond stabilization; a boundary below $S_G$=1 is incompatible with the stated assumptions. A reproducible sub-threshold boundary that survives all stated corrections across independent calibration routes would instead challenge the framework's applicability assumptions themselves. |

Interpretation. A passing result identifies a reproducible interface boundary. Agreement with $S_{G,D}^{eq}$ tests the pure-repeat interpretation; disagreement must be traced to state, calibration, transport, or loss terms. A null result is ambiguous until the miscut control has been run. Near equilibrium, step-flow can self-terminate at positive bulk supersaturation or reservoir driving force once faceting has removed the last step source, so if advance resumes only on the vicinal seed, the earlier null measured the availability of kinks rather than the sign of the thermodynamic driving force.

## S4.3. Protocol 3: isotope-attributed net reservoir-fed overgrowth

**Question.** Did carbon from the external reservoir create new diamond outside the original seed volume?

**Table S7. Protocol 3: isotope-attributed net reservoir-fed overgrowth.**

| Element | Requirement |
|---|---|
| Hypothesis | Carbon supplied through the reservoir forms new, crystallographically continuous diamond outside the original seed volume. |
| Controlled state | Isotopic composition of source, seed, gas, and all carbon-bearing hardware; seed surface map and fiducials; source-seed separation; growth time; cleaning and quench procedure; analysis depth resolution. |
| Primary observable | Depth-resolved isotope fraction in diamond, registered outward interface displacement, crystallographic continuity, bonding analysis, cleaned mass change, and isotope-resolved phase inventory. |
| False positives | Surface residue, trapped particles, isotope exchange without net motion, dissolution-reprecipitation within the original volume, detached diamond, carbide-mediated transfer, or direct gas deposition through a bypass. |
| Required controls | Pre-run seed isotope and topography maps; blank and no-source runs; inverted source/seed isotope contrast; masked or protected reference regions; cross-sectional microscopy; aggressive residue removal; whole-reactor isotope accounting. |
| Quantitative success criterion | The isotope-rich diamond signal exceeds seed and blank controls by at least five standard uncertainties; the enriched layer exceeds three times the depth resolution; outward displacement exceeds three standard uncertainties; continuity and diamond bonding are confirmed; and carbon/isotope balances close within combined uncertainty. |
| Falsifying outcome | Reservoir-fed outward overgrowth is falsified if isotope enrichment lacks registered outward motion, |

| Element | Requirement |
|---|---|
| | crystallographic continuity, or source attribution. Competitor isotope inventories instead reduce or falsify claimed selectivity, dominance, or feed-attributed yield. |

Interpretation. Isotope in diamond is not enough. The new material must also be outside the registered original interface and continuous with the seed.

## S4.4. Protocol 4: reversible finite-graphite boundary

**Question.** Does one controlled finite graphite state lose carbon at a reservoir condition that still advances the diamond seed?

**Table S8. Protocol 4: reversible finite-graphite boundary.**

| Element | Requirement |
|---|---|
| Hypothesis | A specified finite graphite state and path have a reversible carbon-transfer boundary $S_G^{\mathrm{finite,eq},\lambda}(N_C,\chi_G,M)$ that may lie above $S_{G,D}^{\mathrm{eq}}$. |
| Controlled state | Immobilized, narrowly distributed features; layer number or thickness; edge orientation and reconstruction; termination; support; strain; defects; passivation; non-carbon exchange inventories; local temperature and transport field. A matched separate cell is preferred; no macroscopic graphite buffer remains. |
| Primary observable | Short-time initial change in an independent carbon-inventory observable—thickness/volume, calibrated atom count or mass, or isotope-resolved carbon—together with area/perimeter and post-run state analysis versus $S_G$, size, and path. |
| False positives | Reconstruction without carbon transfer, layer-number change, detachment, coalescence, Ostwald ripening, support reaction, passivation, beam damage, and irreversible state drift. |
| Required controls | Multiple sizes and feature populations; repeated bracketing from both sides; initial-rate extrapolation before state drift; post-run edge/support analysis; matched no-diamond and no-graphite cells; perturbation bound if a witness is co-located. |
| Quantitative success criterion | A reversible sign-change locus is resolved for the declared state and path. For the experimentally realized diamond front, a pairwise interval is claimed only when the uncertainty distribution for the difference between $S_G^{\mathrm{finite,eq},\lambda}$ and the measured diamond-front boundary of Protocol 2 is positive after including activity-mapping uncertainty and the chosen high-temperature thermochemical-model uncertainty. If the latter is unavailable, the result is reported as conditional on the stated thermochemical data and model for $S_{G,D}^{\mathrm{eq}}$ rather than as a 95% confidence claim. Comparison of the measured diamond-front boundary with the intrinsic completed-repeat threshold is a separate test of whether the front has reached the pure macroscopic limit. Concurrent diamond advance and finite-graphite carbon loss at the same local $S_G$ is the strongest comparison. |
| Falsifying outcome | The finite-graphite boundary is at or below the measured diamond-front boundary, no state-resolved reversible boundary survives the controls, or the apparent change lacks |

| Element | Requirement |
| --- | --- |
| | an independent carbon-inventory signal. The result then supplies no pairwise thermodynamic interval, although kinetic suppression may remain. |

Interpretation. Geometry change alone is insufficient; the experiment must show reversible carbon transfer for one declared state and path.

On-seed witness variant. Competition at the seed front is one decisive part of a run: a patch nucleated there can spread, be overgrown and enveloped, or redissolve. Registered graphitic, amorphous-carbon, or carbide witnesses can make that local competition measurable, but a fabricated patch only approximates the state the system would nucleate. With precharacterized source-patch isotope contrast and patch-inventory tracking, isotope turnover can distinguish a static patch from active exchange. Envelopment must be normalized to the pre-run patch census and analyzed volume and reported as product quality. It does not measure competitor suppression. Bulk or free-surface competitors, transport, and feed depletion remain separate. Interrupted runs that identify where each non-diamond phase first appears are more representative than a single fabricated witness. Controls must exclude delamination, transfer residue, quench deposition, incomplete recovery, and sampling bias.

## S4.5. Protocol 5: stochastic competitor rates and interrupted-quench inventories

**Question.** How rapidly do graphite, carbide, and amorphous-carbon competitors appear under the same exposure that advances seeded diamond?

**Table S9. Protocol 5: stochastic competitor rates and interrupted-quench inventories.**

| Element | Requirement |
| --- | --- |
| Hypothesis | Seeded diamond can advance while graphite, carbide, and amorphous-carbon rates and retained inventories remain below declared bounds over a stated exposure. |
| Controlled state | Reactor volume and surface area, time-dependent heterogeneous-site population, seed area, validated or model-conditional $S_G$, temperature field, gas and alloy history, run time, quench rate, sampling fraction, detection efficiency, recovery efficiency, and detection limit. |
| Primary observable | Time-resolved counts and sizes of competing nuclei, spatially resolved phase maps, phase-resolved carbon inventories, and seeded-diamond displacement across interrupted runs. |
| False positives | Quench precipitation, cleaning loss, missed sub-detection nuclei, limited sampling, particles imported from upstream, or phase conversion during cooling. |
| Required controls | Multiple interruption times; rapid and alternative quench protocols; blanks without seed or carbon feed; in situ observation where possible; whole-reactor sampling; standards for detection efficiency and phase recovery. |
| Quantitative success criterion | Competitor rates are reported with confidence intervals or, for zero events, the Equation S23 upper bound using the correct time-dependent exposure and quantified detection/recovery efficiency. The event-count model is checked for dispersion; a hierarchical or non-Poisson model is used when independence fails. Seeded diamond has positive registered growth, and competitor rates and retained fractions remain below predeclared bounds. |
| Falsifying outcome | Any competitor exceeds its predeclared rate or retained-inventory bound, including after improved sampling or quench control; or the claimed seeded growth is not registered and source-attributed. |

Interpretation. This protocol measures kinetic selectivity. It does not define the finite-graphite equilibrium boundary of Protocol 4.

## S4.6. Protocol 6: gas-fed thin-film source-to-seed transport

**Question.** Can a finite metal film carry carbon from the gas-facing surface to a buried diamond front without the result being caused by pinholes, direct CVD, or a dominant surface-carbon sink?

**Table S10. Protocol 6: gas-fed thin-film source-to-seed transport.**

| Element | Requirement |
|---|---|
| Hypothesis | A finite metal film transports carbon from a gas-side boundary to a buried diamond front, and the observed rate can be described by coupled gas-side kinetics, film transport, and buried-interface kinetics while free-surface graphite is kinetically suppressed. |
| Controlled state | Film thickness and continuity, alloy composition, temperature profile, gas composition and pressure, free-surface catalyst state, seed orientation and termination, buried-interface area, and run time. |
| Primary observable | Isotope-resolved concentration profiles across the film, free-surface products, buried-interface velocity, film morphology, gas conversion, and complete carbon inventory. |
| False positives | Pinholes or direct gas access to the seed, film dewetting or solidification, lateral transport from an uncontrolled reservoir, free-surface graphite later transferred to the seed, and growth during quench. |
| Required controls | Thickness series; blocked-interface and inert-substrate controls; isotope contrast; cross-sectional imaging before and after; free-surface phase analysis; variation of gas-side transfer and buried-interface kinetics independently where possible. |
| Quantitative success criterion | One physically consistent reaction-diffusion parameter set predicts profiles and rates across multiple thicknesses and boundary conditions within stated uncertainty; isotope-attributed outward diamond growth is registered; and free-surface and buried competitor inventories are bounded. |
| Falsifying outcome | The result is attributable to pinholes or direct CVD, no parameter set can reconcile profiles and rates, or a graphitic/amorphous free-surface layer becomes the dominant carbon sink under the conditions required for buried growth. |

Interpretation. A passing result must explain more than one film thickness and more than one boundary perturbation with the same physical model.

A thick recovered crystal is not the first test of the proposal. The stronger result is agreement among calibrated activity, reversible front motion, isotope attribution, finite-competitor response, kinetic rates, and common-basis inventories.

## S5. BN: pair stoichiometry, B/N imbalance, charge, and motif comparison

### S5.1. Define the three thermodynamic coordinates before drawing a map

**Purpose.** A BN growth environment cannot be described by one elemental chemical potential. Pair incorporation, B-rich versus N-rich imbalance, and electronic charge control are separate coordinates; the third is an electron electrochemical potential rather than a chemical potential.

Choose explicit elemental reference states and define

$$\Delta\mu_B=\mu_B-\mu_B^{\mathrm{ref}}, \qquad \Delta\mu_N=\mu_N-\mu_N^{\mathrm{ref}}. \qquad \text{(S30)}$$

Define the convenience coordinates

$$\Delta\mu_{\mathrm{pair}}=\Delta\mu_B+\Delta\mu_N, \qquad \Delta\mu_{\mathrm{imb}}=\Delta\mu_B-\Delta\mu_N. \qquad \text{(S31)}$$

Define the full conjugate potentials as $\mu_{\mathrm{pair}} = \mu_B + \mu_N$ and $\mu_{\mathrm{imb}} = \mu_B - \mu_N$. Their excess coordinates are given in Equation S31. With $N_{\mathrm{pair}} = (N_B + N_N)/2$ and $N_{\mathrm{imb}} = (N_B - N_N)/2$, $\mu_B\, \mathrm{d}N_B + \mu_N\, \mathrm{d}N_N = \mu_{\mathrm{pair}}\, \mathrm{d}N_{\mathrm{pair}} + \mu_{\mathrm{imb}}\, \mathrm{d}N_{\mathrm{imb}}$. A neutral BN-pair event changes $N_{\mathrm{pair}}$ by one and $N_{\mathrm{imb}}$ by zero. A 1:1 feed fixes the delivered atomic ratio but not the interfacial imbalance.

For an event that transfers one neutral B–N pair and no other species,

$$\Delta Y_{\mathrm{BN,pair}}=G_{\mathrm{front}}(s')-G_{\mathrm{front}}(s)-\mu_B-\mu_N. \qquad \text{(S32)}$$

For charged or redox-coupled events, use the molar electron **electrochemical** potential $\tilde{\mu}_e$:

$$\Delta Y=G_{\mathrm{front}}(s')-G_{\mathrm{front}}(s)-\Sigma_{i\neq e}\ \nu_i\mu_i^{\mathrm{res}}-\nu_e\tilde{\mu}_e. \qquad \text{(S33)}$$

Here $\nu_e > 0$ denotes electrons transferred from an electronic reservoir into the front; the summation in Equation S33 excludes electrons. The electrical reference, charge-compensation or neutrality condition, and any field or capacitance energy must be declared. Periodic charged-supercell energies require finite-size and potential-alignment corrections.[46]

Isotope-labeled seeds could also be valuable for BN studies. Preparation and availability of $^{11}$B-enriched and/or $^{10}$B-enriched cBN seeds would benefit the community by enabling thermodynamic, kinetic, and growth experiments with a feed of contrasting boron-isotope composition. Such experiments could distinguish boron retained in or released from the seed from boron supplied by the reservoir. $^{15}$N labeling should be used in the same way to distinguish nitrogen pathways. A $^{15}$N-enriched nitrogen feed, or eventually a cBN seed with known $^{15}$N enrichment, could distinguish nitrogen supplied by the reservoir from nitrogen released from the seed or from other nitrogen-bearing materials in the apparatus. Boron- or nitrogen-isotope contrast would not by itself establish net outward growth; crystallographic continuity and registered interface motion would still be required.

### S5.2. What a BN compatibility map can—and cannot—establish

**Purpose.** Chemical-potential inequalities can identify allowed compositions and defect conditions. They do not by themselves select cBN or wBN over layered BN. Here cBN is cubic BN, wBN is wurtzite BN, and layered BN denotes the $sp^2$-bonded hexagonal, rhombohedral, and turbostratic variants collectively.

Competing phases impose inequalities in a declared $(\Delta\mu_B,\Delta\mu_N,\tilde{\mu}_e)$ slice. These inequalities may identify conditions compatible with stoichiometric pair incorporation and selected defect populations. A map of those conditions is a **compatibility map**, not a measured bonding-motif phase diagram.

The distinction matters because high-level calculations disagree on BN polymorph ordering. A random-phase-approximation study reports cBN as the 0 K ground state with hBN entropically stabilized above approximately ambient temperature, whereas fixed-node diffusion Monte Carlo reports hBN as lower at 0 and 300 K.[47,48] Metal- and metal-nitride-solvent routes have produced cBN at high pressure and high-quality hBN from molten iron at atmospheric pressure.[49,50] The former study also reports that the iron-group catalysts — iron, nickel and manganese — yielded only hexagonal BN even at 100 000 atm, and the latter was run in a hydrogen-bearing atmosphere at approximately 1 atm and produced no cubic phase. In situ hBN microscopy and first-principles hydrogen-dependent nucleation studies identify effects of steps, terraces, termination, metal environment, and hydrogen.[51,52] Neither study establishes a low-pressure window in which cBN is favored at growth-relevant temperatures; the random-phase-approximation work places the ambient-pressure cubic-to-hexagonal crossover at 335 ± 30 K.

A credible low-pressure BN test must therefore report the reference states; $\Delta\mu_B$, $\Delta\mu_N$, and $\tilde{\mu}_e$ or the electrical boundary condition; the actual transferred growth unit; phase-resolved B and N inventories; defect and polarity controls; and either reversible feature response or rate-resolved competition between tetrahedral and layered BN states. Pair and defect compatibility are necessary but not sufficient for motif selection.

# S6. A thin metal film is a two-boundary reaction-diffusion system

## S6.1. Conserve carbon before choosing a transport approximation

**Purpose.** A thin film may relax rapidly across its thickness and still be limited by gas-side uptake, buried-interface reaction, or coupled multicomponent transport.

Let $z=0$ denote the gas/$M$ surface and $z=h$ the buried $M$/diamond interface for the chosen film-forming metal or alloy $M$. For an effective dissolved-carbon concentration $c_C(z,t)$,

$$(\partial c_C)/(\partial t)=-(\partial j_C)/(\partial z)+R_C^{\mathrm{bulk}}. \qquad \text{(S34)}$$

$R_C^{\mathrm{bulk}}$ is a signed volumetric source term: it is positive for net production of dissolved carbon and negative for net consumption by homogeneous reactions such as carbide or cluster formation. Equation S34 is a conservation law; it does not specify the transport mechanism.

In a multicomponent nonideal liquid, flux is generally driven by coupled chemical-potential gradients and thermal forces. A linearized one-dimensional form is

$$j_i=-\Sigma_j\, L_{ij}(\partial)/(\partial z)((\mu_j)/(T))+L_{iT}(\partial)/(\partial z)((1)/(T)). \qquad \text{(S35)}$$

Equation S35 is a linear irreversible-thermodynamic form valid near local equilibrium.[3] The coefficients require a stated independent-component set, concentration and flux basis, stationary reference frame, and positive-semidefinite response matrix. Cross-coefficients and thermal forces can drive carbon against its own $\mu_C$ gradient. Monotonic $\mu_C$ may be inferred only for an isothermal diagonal response with positive $L_{CC}$ in the stated frame and no omitted advection, electromigration, body force, or side sink. Fickian diffusion is a special case.

## S6.2. Give each boundary its own kinetics

**Purpose.** Carbon entering at the free surface, crossing the film, and reacting at the buried boundary are distinct processes and must not be collapsed into one fitted rate constant.

Take positive $j_C$ to point toward the buried diamond interface. At the gas-facing boundary,

$$j_C(0)=R_{g\to L}-R_{L\to g}-R_{C,solid,surface}-R_{other,surface}. \qquad \text{(S36)}$$

All rates are nonnegative magnitudes on the same molar-carbon, area, and time basis. The aggregate rate $R_{C,solid,surface}$ includes free-surface graphite and amorphous-carbon formation; the other terms represent gas uptake, desorption, and other free-surface sinks.

At the buried boundary,

$$j_C(h)=(j_{D,k}^{\mathrm{att}}-j_{D,k}^{\mathrm{det}})+j_{\mathrm{buried,cons}}-j_{\mathrm{buried,prod}}. \qquad \text{(S37)}$$

Here $j_{\mathrm{buried,cons}}>0$ is carbon consumed by buried side reactions such as carbide or graphite formation, and $j_{\mathrm{buried,prod}}>0$ is carbon released into the liquid by dissolution or decomposition of buried phases. Passivation that only blocks sites belongs in the boundary kinetic laws and is not a carbon sink unless the passivating reaction itself consumes carbon.

Equations S34–S37 require gas conversion, free-surface carbon, dissolved inventory, buried products, and outflow to be included in one balance.

For a linear single-component diagnostic,

$$\tau_{\mathrm{diff}}\sim(h^2)/(D_{\mathrm{eff}}), \qquad \mathrm{Da}_{\mathrm{gas}}=(k_{\mathrm{gas}}h)/(D_{\mathrm{eff}}), \qquad \mathrm{Da}_{\mathrm{buried}}=(k_{\mathrm{buried}}h)/(D_{\mathrm{eff}}). \qquad \text{(S38)}$$

$\tau_{diff} \sim h^2/D_{eff}$ is the characteristic diffusion-time scale across a film of thickness $h$; the exact relaxation time depends on geometry and boundary conditions. $k_{gas}$ and $k_{buried}$ are first-order interfacial mass-transfer coefficients in m s$^{-1}$. The Damköhler numbers compare each boundary-reaction scale with diffusion across the film. They are diagnostics only, and neither one proves which step limits the rate.

In the restricted passive, isothermal diagonal-response limit, with positive $L_{CC}$ in the stated stationary frame, no advection or body-force term, and no side sink between the boundaries, carbon flux that feeds a buried macroscopic diamond front implies that the supplying free-surface carbon potential is not lower than the buried-interface potential. For a pure, unstressed repeat front requiring $S_G > S_{G,D}^{eq} > 1$, the supplying surface is then above graphite saturation. In the general multicomponent nonisothermal problem, Equation S35 must be solved before inferring the $\mu_C$ or $S_G$ profile.

Suppression of free-surface graphite or amorphous carbon is therefore kinetic only in that restricted case. SiC studies establish metal-alloy-flux-assisted growth and thin precursor-layer processing; they do not establish a stable continuous thin-liquid reservoir over the growth area or diamond-over-graphite selection.[53,54] Any thin-film candidate requires separate validation of wetting, continuity, liquid stability, evaporation and composition drift, gas-side chemistry, and direct gas access to the seed.

## S6.3. Use several perturbations to identify the controlling resistance

**Table S11. Expected responses to perturbations under idealized limiting cases; no single response identifies the controlling resistance by itself.**

| Perturbation | Transport-limited signature | Gas-side-limited signature | Buried-interface-limited signature |
|---|---|---|---|
| Increase film thickness | Rate decreases and concentration gradient grows | Weak effect after surface state is fixed | Weak effect while diffusion remains fast |
| Increase gas-side transfer or precursor activity | Rate increases only until film or buried-interface limitation is reached | Strong response | Saturating response |
| Change seed miscut or termination | Weak direct effect | Weak direct effect | Strong response through attachment and passivation |
| Change stirring or imposed gradient | Strong response if external or internal transport controls | May alter surface supply | Weak after local chemical potential is fixed |
| Apply an isotope pulse | Broad, delayed arrival through the film | Delay at uptake boundary | Accumulation in the film followed by slow incorporation |

## S6.4. Measurement primitives already available in liquid-phase epitaxy and vapor–liquid–solid growth

Liquid-phase epitaxy and vapor–liquid–solid growth provide relevant precedents for condensed reservoirs, seeded growth, transport control, and phase-selective masking.[38,39] Their demonstrated measurements are separated below from adaptations proposed for the present framework.

Konuma documents near-equilibrium seeded growth, step-flow behavior, self-termination after step-source exhaustion, and suppression of growth on amorphous or oxide-masked regions.[39] The same apparatus could be adapted to approach a boundary from meltback and growth sides, and stirring at independently fixed driving force could test transport versus interface limitation; these are diagnostics I propose, and that source reports neither of them. Solvent inclusions can mark geometric trapping during facet coalescence, but they are not by themselves a measure of exhausted kink supply. Masked-region results isolate seeded growth relative to those regions. Nucleation elsewhere is not excluded.

A verified thin-liquid precedent is available: Demuth et al. grew epitaxial Ge on Si electrochemically through an approximately 8–10 µm eutectic Ga-In film at 90 °C and ambient pressure.[55] This demonstrates a confined liquid-metal geometry and seeded crystallographic outcome. It does not establish kink-level selection or transfer its boundary conditions to carbon.

The vapor–liquid–solid literature extends metal-alloy-flux-assisted SiC growth to Pt-Si melts and pre-deposited $NiSi_2$ layers.[53,54] These studies establish alloy-assisted growth and thin as-deposited alloy-layer processing. Neither shows a stable continuous liquid film across the seed. The same group later resolved step advance, recession, and renewed advance at a SiC solution interface in a Si-Cr solvent, but during an uncontrolled cooling transient and without a calibrated carbon potential.[41] No study in this literature combines graphite calibration, reversible bracketing, and transport diagnostics in one geometry.

One diagnostic can be transferred. In a Si-Ni solvent the step-advance rate divided by the step distance scaled as the inverse of the step height over a range of about 10 to 58 nm, which identifies a diffusion-controlled regime. The fitted coefficient was constant within error at 1620, 1655, and 1683 °C.[40] Adding Al removed that dependence, but the step heights then spanned only about 14 to 32 nm, so the second result is the weaker of the two. Applied to a diamond front, the same comparison would separate transport control from attachment control without fitting an overall rate constant, provided step heights can be resolved over a comparable range.

A schematic comparison by Capper et al. places typical liquid-phase epitaxy near 0.1–1% supersaturation and chemical vapor deposition near 10–100%.[38] Because the fields may use different growth units and reference states, these values are order-of-magnitude ranges rather than one calibrated activity scale. If a candidate metal flux is experimentally shown to operate near equilibrium, a favorable sign can still require a sufficient population of active steps and kinks. The operating regime has to be measured. Analogy will not assign it.

A model is credible only if one parameter set explains several independent perturbations and closes both carbon and isotope balances. Fitting one growth rate with many unconstrained parameters is not validation.

## S6.5. Seeded nitride flux growth supplies process tests, not bonding-network selection

Aoki et al. mapped three outcomes for GaN seeds in Na-Ga flux in temperature–$N_2$-pressure space, for runs of 50 to 500 h with duration not resolved as a variable: seed decomposition, seed-only growth, and seed growth accompanied by off-seed nucleation; growth also depended strongly on which polar face contacted the melt.[56] Morishita et al. then observed a nonmonotonic seeded rate together with GaN nucleation near the gas-liquid interface and attributed the decline to nitrogen consumption by the off-seed crystals and reduction of the supplying gas-liquid area.[57] These are strong precedents for separately scoring seed loss, seed gain, off-seed nuclei, position, polarity, and time. Their reported boundaries and supersaturation values depend on the protocol and the model used. They are not thermodynamic phase boundaries. The supersaturation axis of that work is a computed upper bound that its authors describe as exceeding the true value, so the precedent shares the calibration limitation and does not resolve it.

An early Ga-Na study combined a GaN dissolution curve with the nitrogen pressure at which GaN was first recovered and an assumed Sieverts' relation to infer dissolved nitrogen.[58] Kawamura et al. used a 6 µm GaN film for the dissolution measurement. Morishita et al. later found a substantially lower GaN solubility after increasing the melt mass about 26-fold, using crystals larger than 100 µm instead of a nonuniform film, continuously shaking the melt, quenching it, and determining dissolution by gravimetric recovery.[59] In the revised measurements, Li or Ca increased both measured GaN solubility and inferred nitrogen solubility; the inferred nitrogen increase was larger under the tested conditions. The reported "absolute value of supersaturation" was an atomic-percent composition difference, not an activity ratio, ln *S*, a chemical-potential difference, or a reversible seed-front affinity. The measurement is a useful precedent for how to calibrate. Nothing in it transfers to a carbon activity model. Dissolved nitrogen was inferred rather than measured directly, the pressure mapping assumed Sieverts' relation, and GaN recovery after 96 h supplied the threshold. The proposed Na-driven ionization and Li/Ca-binding mechanisms were not observed directly.

Flux-film-coated GaN growth adds geometry and history. The original process repeatedly coated a seed with Ga-Na solution and demonstrated epitaxial growth at 800 °C and 9.5 atm $N_2$.[60] Later work showed that residual film retention can depend on surface morphology and that pool-to-film cycling can suppress giant steps and melt inclusions while changing Al, C, Si, and Mn incorporation.[61,62] A 2026 comparison reduced off-seed polycrystalline GaN from about 11% to about 1% of the charged gallium at 900 °C/5 MPa relative to 870 °C/3 MPa at similar total yield; temperature and pressure changed together, the two conditions are single runs of a multi-stage process on a multi-point-seed substrate with carbon added to the melt, and supersaturation was inferred.[63] Thus film thickness, wetting, replenishment, gas exposure, step topology, impurity reservoirs, and product partition must be measured as coupled controls.

An AlN solution-growth study used a Cr-Co-Al solvent and a temperature gradient under 1 atm $N_2$ to obtain a rough, approximately 300 µm free-standing layer in 1 h, from which the authors inferred, but did not measure, concurrent dissolution of the sapphire substrate; the reported mean thickening rate was $247 \pm 51$ µm $h^{-1}$.[64] This supports spatial reservoir design—making the gas-facing region less favorable for parasitic nucleation while maintaining growth at the substrate—but it does not measure an interface rate, establish the proposed microscopic mechanism, or address competing $sp^2/sp^3$ networks. The changing substrate geometry belongs in the material balance.

# S7. Compact symbol and reporting checklist

## S7.1. Core symbols used in both documents

**Table S12. Core symbols.**

| Symbol | Meaning | Unit |
|---|---|---|
| $s, s'$ | Declared constrained-equilibrium or metastable front basins | — |
| $M$ | Chosen metal or alloy that constitutes the carbon-bearing flux; local composition and reservoir/interfacial state are separate quantities | — |
| $Y, \Delta Y$ | Reservoir-transformed state function and event change | J per mol of identical declared events |
| $\nu_i$ | Net reservoir-to-front transfer of species *i* per mole of events | mol species *i* per mol event |
| $\mu_i^{\mathrm{res}}$ | Reservoir chemical potential | J mol$^{-1}$ species |
| $\mu_C^L$ | Carbon chemical potential in the chosen metal-flux reservoir | J mol$^{-1}$ C |
| $\mu_u^{\mathrm{res}}$ | Stoichiometric reservoir potential of one growth unit | J mol$^{-1}$ growth units |
| $\mu_{k,\varphi}^{\mathrm{rep}}$ | Modern completed translational repeat potential | J mol$^{-1}$ growth units |
| $g_\varphi^{\mathrm{bulk}}$ | Bulk Gibbs free energy per mole of growth units | J mol$^{-1}$ growth units |
| $\Delta G_{nonrep}$ | Remainder after subtracting congruent repeats | J mol$^{-1}$ event basis |
| $S_G, S_{G,D}^{\mathrm{eq}}$ | Graphite-normalized carbon activity and its pure-diamond equilibrium threshold | Dimensionless |
| $\mu_{C,G}^{\mathrm{finite},\lambda}$ | Marginal carbon-transfer cost of a declared finite graphite state and path | J mol$^{-1}$ C |
| $\beta, \tilde{\beta}$ | Edge line free energy and step stiffness | J m$^{-1}$ |
| $\gamma$ | Interfacial free energy | J m$^{-2}$ |
| $\Gamma_C$ | Carbon activity coefficient under the stated standard-state convention | Dimensionless |
| $\bar{A}_u, \bar{V}_u$ | Molar area and volume per mole of growth units | m$^2$ mol$^{-1}$, m$^3$ mol$^{-1}$ |
| $A_{int}$ | Retained-flux interfacial area used for rate normalization | m$^2$ |
| $j$ | Molar interfacial or transport flux on a declared area basis | mol growth units m$^{-2}$ s$^{-1}$ unless another basis is stated |
| $\mathcal{J}_\alpha$ | Activated-event or nucleation rate density | Events per declared exposure basis |
| $\Delta Y^\ddagger$ | Molar activation free energy | J mol$^{-1}$ |
| $\Delta\mu_{\mathrm{pair}}, \Delta\mu_{\mathrm{imb}}, \tilde{\mu}_e$ | BN pair, imbalance, and electron electrochemical coordinates | J mol$^{-1}$ on the explicitly declared BN-pair, atomic, or electron basis |
| $\mathcal{A}$ | Thermodynamic affinity of the declared forward event, $\mathcal{A} \equiv -\Delta Y$ | J per mol of identical declared events |
| $b_i$ | Stoichiometric coefficient of species i | mol species i per mol growth units |

| Symbol | Meaning | Unit |
|---|---|---|
| | in one growth unit | |
| $\Delta g_{DG}$ | Bulk diamond-minus-graphite molar Gibbs free-energy difference | J mol$^{-1}$ C |
| $\chi_G$, $\lambda$ | Declared finite-graphite state and its growth/dissolution path | — |
| $S_G^{op}$ | Operating carbon activity on the graphite-normalized scale | Dimensionless |
| $H_\alpha$, $\Pr_\alpha$, $\mathcal{E}$, $\varepsilon_\alpha$ | Integrated detected-event hazard, probability of at least one detected event, active exposure, and detection/recovery efficiency | $H_\alpha$, $\Pr_\alpha$, and $\varepsilon_\alpha$ are dimensionless; $\mathcal{E}$ has the declared exposure basis |
| $A$, $\mathcal{P}$, $\ell$; $A_C$, $\bar{A}_C$ | Feature area, perimeter, A/$\mathcal{P}$, basal area per C atom, and its molar value | m$^2$; m; m; m$^2$ per C atom; m$^2$ mol$^{-1}$ C |

## S7.2. Origin and role of the equation groups

The equations in this SI do not all have the same status. The table below distinguishes definitions and accounting identities introduced here from standard relations taken from thermodynamics, kinetics, and transport theory, and from numerical constructions used only for the present estimates. This prevents an identity from being mistaken for a new physical law and shows which parts require experimental validation.

**Table S13. Origin and role of equation groups.**

| Equations | Status | Role and limitation |
|---|---|---|
| S1–S3, S5–S8, S10–S12, S15–S18, S24–S33 | Definitions, exact bookkeeping identities, or author-chosen decompositions | Organize a declared event, inventory, coordinate system, or source-attributed yield. S25 is exact only for a complete control volume and consistent rate basis; S27 requires source attribution or a sole-source restriction. |
| S4 | Modern local-detailed-balance relation; historically related to BCF detailed balancing of equilibrium kink configurations | Connects conditional elementary transition rates between declared basins. It is not attributed as a formula of the early sources. A coarse-grained flux ratio additionally requires consistent state populations, activities, and no hidden cycle. |
| S9 | Standard activity expression for chemical potential | Converts a declared activity scale into chemical potential. The standard state and activity model must be specified and calibrated. |
| S13–S14 | Author-applied thermochemical construction with standard uncertainty propagation | Gives the working estimate of the diamond threshold near 1300 K. It is conditional on the chosen thermochemical anchor and temperature model. No universal constant is implied. |
| S19 | Author-applied marginal derivative for a self-similar two-dimensional feature using the corrected graphite basal area | Corrects the earlier mean-excess estimate by differentiating edge energy along a declared growth path. The illustrative numerical scale is not a universal graphite crossover. |
| S20 | Standard isotropic Gibbs-Thomson curvature relation | Gives a curvature contribution under isotropic, equilibrium assumptions; |

| Equations | Status | Role and limitation |
|---|---|---|
| | | anisotropy, faceting, reconstruction, and support interactions require a more specific treatment. |
| S21 | Standard activated-rate form | Separates barrier and prefactor from thermodynamic direction. A fitted rate does not identify the microscopic mechanism without independent constraints. |
| S22–S23 | Standard integrated-hazard and Poisson-event relations | Convert event counts into stochastic bounds under the stated independent-event or hazard model. S23 defines a conventional one-sided 95% upper limit; heterogeneous exposures can require a hierarchical or non-Poisson model. |
| S34–S37 | Conservation laws and linear irreversible-thermodynamic transport form | Enforce bookkeeping across the film. The linear form applies near local equilibrium in a stated frame; monotonic-potential inference additionally requires positive diagonal response and no omitted advection, body force, or side sink. |
| S38 | Standard diffusion-time and Damköhler diagnostics | Characteristic diffusion-time and Damköhler diagnostics. $\tau_{diff} \sim h^2/D_{eff}$ is a scaling estimate whose numerical factor depends on geometry and boundary conditions. |

### S7.3. What must accompany any claimed boundary or operating window

A reported thermodynamic boundary or operating window should state the applicable items below. Isotope inventories are required when provenance or feed-attributed yield is claimed; inclusion metrics are required when overgrowth quality or selectivity is claimed.

- the initial and final states and the transferred growth unit;
- all independent chemical reservoirs and the activity standard state;
- temperature, pressure, mechanical, and electrical controls;
- front geometry, stress, curvature, composition, termination, and passivation;
- whether the result is a completed-repeat boundary, a pairwise finite-object marginal boundary for a declared path, an apparent zero-velocity locus, or a stochastic rate threshold;
- the approach direction and reversibility test;
- transport and loss corrections;
- uncertainty, observation time, time-dependent active area or volume, detection efficiency, recovery efficiency, and detection limit;
- phase-resolved inventories on a common basis and, when isotope labeling or source attribution is used, isotope-resolved inventories on that basis;
- for recovered overgrowth, a non-diamond inclusion census—the number density, size distribution, and, where possible, phase assignment—with envelopment reported as product quality rather than competitor suppression;
- and the observation that would falsify the interpretation.

If the claim-relevant state, reversibility, transport/loss, or uncertainty requirements are not satisfied, report an apparent or protocol-dependent onset rather than a thermodynamic boundary.

## References

References in this Supporting Information are numbered independently of those in the main text.